\documentclass{article}
\usepackage[utf8]{inputenc}
\usepackage[in]{fullpage}
\usepackage[affil-it]{authblk}
\usepackage[numbers]{natbib}
\usepackage{multirow}%
\usepackage{amsmath,amssymb,amsfonts}%
\usepackage{amsthm}%
\usepackage{mathrsfs}%
\usepackage[title]{appendix}%
\usepackage{xcolor}%
\usepackage{textcomp}%
\usepackage{manyfoot}%
\usepackage{booktabs}%
\usepackage{algorithm}%
\usepackage{algorithmicx}%
\usepackage{algpseudocode}%
\usepackage{listings}%
\usepackage{natbib}
\usepackage{siunitx}
\usepackage{enumerate}
\usepackage{graphicx}
\usepackage{url}

\title{Variable Selection Using Nearest Neighbor Gaussian Processes}

\author[1,3]{Konstantin Posch}
\author[1,3]{Maximilian Arbeiter}
\author[2]{Christian Truden}
\author[4]{Martin Pleschberger}
\author[1]{Jürgen Pilz}

\affil[1]{Department of Statistics, University of Klagenfurt, Universit\"atsstraße 65-67, 9020-Klagenfurt, Austria}
\affil[2]{Department of Operations, Energy and Environmental Management,
University of Klagenfurt, 9020-Klagenfurt, Austria}

\affil[3]{Infineon Technologies Austria AG,
Siemensstraße 2, 9500-Villach, Austria}

\affil[4]{KAI Kompetenzzentrum Automobil- und Industrieelektronik GmbH,
Technologiepark Villach Europastraße 8,
9524-Villach, Austria}

\begin{document}

\maketitle

\begin{abstract}
We introduce a novel Bayesian approach for variable selection using Gaussian process regression, which is crucial for enhancing interpretability and model regularization.
Our method employs nearest neighbor Gaussian processes, serving as scalable approximations of classical Gaussian processes.
Variable selection is achieved by conditioning the process mean and covariance function on a random set that represents the indices of contributing variables. A  priori beliefs regarding this set control the variable selection, while reference priors are assigned to the remaining model parameters, ensuring numerical robustness in the process covariance matrix. We propose a Metropolis-Within-Gibbs algorithm for model inference. Evaluation using simulated data, a computer experiment approximation, and two real-world data sets demonstrate the effectiveness of our approach.
\end{abstract}

\noindent%

\section{Introduction}
\label{sec:intro}
Due to the omnipresent and ongoing digitalization process, more and more data is collected by companies, governments, and organizations of all kinds. Consequently, the urge to analyze dependencies between observed variables is also growing. Regression techniques are often the methods of choice for this task. The application of regression models is straightforward if there is a large number $n$ of observations of a comparatively small number $d$ of observed variables.
However, often this does not hold true, and things get challenging. Models start to over-fit since the model complexity cannot be supported by the available sample size \cite{Buehlmann}. 
Moreover, it becomes challenging to identify the most influential variables. For this reason, approaches that automatically detect the most important variables gain increasing attention in many fields, including genetics \citep{GeneSelection}, astronomy \citep{Zheng2007}, and economics \citep{Foster2004}.
In semiconductor manufacturing, there is a vast  number of parameters that are measured during the production process.
Variable selection models help pinpoint specific production parameters that significantly impact the quality of the final product. 

This study introduces a new Bayesian approach to solving the variable selection problem. The novelty of this work is the combination of nearest neighbor Gaussian processes (NNGPs), reference priors, and  a random set $\mathcal{A}$ that specifies the variables that contribute to the regression model.

For the model inference, we propose a Metropolis-within-Gibbs algorithm to sample from the posterior 
distribution. 
In particular, the proposed Markov chain Monte Carlo (MCMC) algorithm makes use of Hamiltonian Monte Carlo (HMC) \citep{DUANE1987}.
HMC is a family of  MCMC methods based on the concept of dynamic systems in physics.
Virtual dynamical systems are created by augmenting hyperparameters with momentum variables, and in the end, a sampling from the distribution of the combined system is performed.
In a Bayesian context, HMC uses the derivatives of the posterior density to generate efficient proposals. In many applications, HMC converges significantly faster than conventional approaches \cite{ DUANE1987, Neal1997, Rasmussen2005}.

The remainder of the work is structured as follows:
Section \ref{sec:rel} reviews related work. 
Following that, Section \ref{sec:methodology} presents our variable selection approach. 
In Section \ref{sec:eval}, we assess the predictive performance of our approach using various datasets.
Lastly, Section \ref{sec:conclusion} summarizes our achievements and concludes the paper.

\section{Related Work}
\label{sec:rel}

If the dependencies of interest are purely linear, simple linear regression models are sufficient for the analysis. There are many well-established approaches to perform variable selection in linear regression models \cite{Tibshirani1996,Fan2001,Hastie2005,Zou2006,Park2008,10.1080/01621459.2013.869223,Alhamzawi2018,POSCH2020}.

Bhattacharya et al. \cite{doi:10.1080/01621459.2014.960967}
introduce and analyze the properties of Dirichlet–Laplace priors, emphasizing their utility in achieving optimal shrinkage in statistical modeling, especially in high-dimensional settings.
Bhadra et al. \cite{10.1214/16-BA1028} tackle the challenges associated with estimating ultra-sparse signals, where the signals of interest are extremely sparse among a vast amount of noise or irrelevant information.
They introduced the Horseshoe+ estimator, which extends the well-known Horseshoe prior by incorporating additional features to enhance its performance in accurately identifying and estimating ultra-sparse signals further.
Chen and Walker \cite{10.1214/18-EJS1529}
 introduce a Bayesian approach to variable selection in high-dimensional linear models using Marginal Solo Spike and Slab priors. These priors offer a computationally efficient way to perform Bayesian variable selection.

However, linearity does not hold in many practical applications, and more sophisticated methods are required. Recently, Gaussian process regression models (GPRMs), originally proposed by O’Hagan \cite{Hagan1978}, gained much attention. GPRMs can be considered an alternative to neural networks since a large class of neural network-based regression models converges to a Gaussian Process (GP) in the limit of an infinite network \cite{Neal1996}. Alternatively, GPs can also be viewed from the perspective of non-parametric Bayesian regression \cite{Rasmussen2005}. Here, Gaussian priors are directly assigned to the space of regression functions. Many empirical studies confirm that GPRMs perform very well compared to other non-linear models \cite{CHEN2007,Jin2008}.
GP models are widely used in spatial statistics, see e.g. \cite{GELFAND201686}.
Moreover, GP regression has become a prevalent approach for the approximation of time-expensive computer models and Bayesian optimization, e.g., \cite{Santner2003,Gramacy2020}.
A survey on high-dimensional GP modeling with applications to Bayesian optimization is given in \cite{Binois2022}.
Advancements to typical GP models allow e.g. for non-stationary flexibility (\cite{sauer2023nonstationary}).
Alternatively, divide-and-conquer GPs such as treed GPs and local approximate GPs avoid computational bottlenecks associated with fitting ordinary GPs on large training data sets (\cite{GramacyLee2008, GramacyApley2015}).
Finally, Deep GPs combine non-stationary and global modeling (\cite{Damianou2013}). They come, however, at the cost of additional computational expenses.

A large class of GPRMs automatically provides some information regarding the importance of the individual variables the model is fitted with. This property is known as automatic relevance determination \citep{Rasmussen2005}. However, with limited training data, such approaches perform poorly, resulting in a high generalization error. One possibility to overcome this problem is adding a penalty term that encourages model fits, where only some variables are considered important. Yi et al. \cite{Yi2011} and Wu et al. \cite{Wu2017} proposed such approaches. 
\cite{Linkletter2006} proposed a related Bayesian approach, which uses a spike and slab prior for the variable selection task. Chen and Wang \cite{CHEN2010} proposed a Bayesian approach that uses a random indicator vector to specify which variables contribute to the model.
Zhang et al. \cite{ZHANG2023107757} use an indicator-based Bayesian variable selection approach for Gaussian process models in computer experiments.

A possible weakness of GPRMs is that the estimation process becomes numerically unstable. To address this issue 
Gu et al. \cite{gu2018} proposed the concept of robust estimation. In a Bayesian manner, robust estimates can be achieved by assigning so-called reference priors to the GP parameters. These priors are recommended by authors of references \cite{Berger2001,paulo2005,KAZIANKA2012, VOLLERT2019}.
As we follow these recommendations, our reference prior assures robust correlation matrices, avoiding the problems arising from near-singularity or near-diagonality of these matrices. 
This also provides a safeguard against effects associated with Jeffreys paradox, which is known to happen in linear regression models.
Another disadvantage of GPRMs is that they are computationally expensive. 
Recently, \cite{Datta2016} introduced so-called nearest neighbor Gaussian processes (NNGPs). These can be considered highly scalable approximations derived from classical GPs.

Variable selection techniques are often applied on spatial data.
\citet{https://doi.org/10.1111/biom.12254} propose a variable selection technique for spatial data that accommodates both local and global variable selection.
For each location, regression coefficients come from a mixture model with the flavor of stochastic search variable selection, but utilize a copula to share information about variable inclusion and effect magnitude across locations.
 The model is applied to identify particularly harmful components of particulate matter.
The work by \citet{10.1093/biomet/asx075} is concerned with spatial variable selection for scalar-on-image regression. A soft-thresholded Gaussian process provides large prior support over the class of piecewise-smooth, sparse, and continuous spatially varying regression coefficient functions.

\section{Methodology}
\label{sec:methodology}

In this section, we define our variable selection approach.
It combines nearest neighbor Gaussian processes (NNGPs) and reference priors.
The variable selection is then based on a random set $\mathcal{A}$ that specifies the variables that contribute to the regression model.

First,  in Section \ref{ss:NNGP}, we briefly reconsider nearest neighbor Gaussian processes.
Then we define the proposed  reference prior in Section \ref{ss:refPrior}.
In Section  \ref{ss:vsm}, we introduce the actual 
variable selection model, followed by the model inference in Section
\ref{ss:inference}.
Finally, in Section \ref{ss:prediction}, we show how predictions can be obtained from the model.

\subsection{Nearest Neighbor Gaussian Process}
\label{ss:NNGP}

A stochastic process $\left\{y(\mathbf{x});\mathbf{x}\in X\subseteq\mathbb{R}^d\right\}$ is called a Gaussian Process (GP) iff for every finite set of indices $\left\{\mathbf x_1,\ldots,\mathbf x_k\right\}\subseteq X$ the random vector $y_{\mathbf x_1,\ldots,\mathbf x_k}=(y(
\mathbf x_1),\ldots,y(\mathbf x_k))^T$ is multivariate normally distributed. 
A GP is fully determined by its mean function $\mu:X\rightarrow  \mathbb{R},~\mathbf{x}\mapsto\mathbb{E}(y(\mathbf{x}))$ and
its covariance function $C:X\times X\rightarrow  \mathbb{R},~(\mathbf{x}_1,\mathbf{x}_2) \mapsto \operatorname{Cov}(y(\mathbf{x}_1),y(\mathbf{x}_2))$. 
In nonparametric Bayesian regression, GPs are often used as prior distribution for the unknown regression function. 
Hence, performing inference in such a model is therefore reduced to the estimation of some unknown parameters used to define the mean function $\mu(\cdot)$ and the covariance function $C(\cdot,\cdot)$.

In the context of the regressions, the variables $\mathbf x_i$ are also called covariates (or predictors).
Let $D=\{(\mathbf{x}_i,y(\mathbf{x}_i))|i=1,\ldots,n\}$ be a finite realization of a GP, i.e., some training data. 
Further, let $S = \{\mathbf{x}_1,\ldots,\mathbf{x}_n\}$ denote the set of covariate constellations for which observations are available.
Then,  $\mathbf C_{S}=[C(\mathbf{x}_i,\mathbf{x}_j)]_{\mathbf{x}_i,\mathbf{x}_j\in S}$ denotes the corresponding covariance matrix. 
Due to the form of the Gaussian density, the model inference is based on the inverse matrix $\mathbf C_{S}^{-1}$ and the determinant $|\mathbf C_{S}|$.

When dealing with large values of $n$, employing GP regression models poses challenges due to the computational expense involved in computing inverse matrices and matrix determinants.
To overcome these issues,
Datta et al. 
 \cite{Datta2016} introduce nearest neighbor Gaussian processes (NNGPs) as highly scalable approximations derived from classical GP models. 
The starting point of an NNGP is a zero mean GP, i.e., $\mu(\cdot)\equiv 0$. 
Then, the main idea behind NNGPs is to write the joint density of $\mathbf{y}_{S}=(y(\mathbf{x}_1),\ldots,y(\mathbf{x}_n))^T$ as a product of conditional densities, using the multiplication theorem, and subsequently replacing the conditioning sets by smaller ones of size at most $m < n$. 
Formally, this is expressed as:
\begin{align}
p(\mathbf{y}_{S})&=p(y(\mathbf{x}_1))\prod\limits_{i=2}^n p(y(\mathbf{x}_i)|y(\mathbf{x}_{i-1}),\ldots,y(\mathbf{x}_1)),\nonumber\\
\tilde{p}(\mathbf{y}_{S})&=\prod\limits_{i=1}^n p(y(\mathbf{x}_i)|y_{N(\mathbf{x}_i)}) ,\label{refDensity}
\end{align}
where $y_{N(\mathbf{x}_i)}$ denotes the vector formed by stacking all the $y(\mathbf{x})$ for which $\mathbf x$ lies in the so-called neighbor set of  $\mathbf{x}_i$, i.e., $N(\mathbf{x}_i)\subset S\setminus\{\mathbf{x}_i\}$, $|N(\mathbf{x}_i)| \leq m$.

By defining the neighbor set $N(\mathbf{x}_i)$ to be any subset of $\{\mathbf{x}_1,\ldots,\mathbf{x}_{i-1}\}$ one can ensure that \eqref{refDensity} is a valid probability density, see \cite{Datta2016}. 
In this work, we follow the approach of  Vecchia \cite{Vecchia1988}, who specified $N(\mathbf{x}_i)$ to be the $m$ nearest neighbors of $\mathbf{x}_i$ among $\{\mathbf{x}_1,\ldots,\mathbf{x}_{i-1}\}$ with respect to the Euclidean distance.
In that case, the density $\tilde{p}(\mathbf{y}_{S})$  is given by:
\begin{align}
\tilde{p}(\mathbf{y}_{S})=\prod\limits_{i=1}^n\mathcal{N}(y(\mathbf{x}_i)|\mathbf{B}_{\mathbf{x}_i}y_{N(\mathbf{x}_i)},F_{\mathbf{x}_i})\label{refDensity2},
\end{align}
where $\mathbf{B}_{\mathbf{x}_i}=\mathbf{C}_{\mathbf{x}_i,N(\mathbf{x}_i)}\mathbf{C}_{N(\mathbf{x}_i)}^{-1}$ and $F_{\mathbf{x}_i}=\mathbf{C}(\mathbf{x}_i,\mathbf{x}_i)-\mathbf{C}_{\mathbf{x}_i,N(\mathbf{x}_i)}\mathbf{C}_{N(\mathbf{x}_i)}^{-1}\mathbf{C}_{N(\mathbf{x}_i),\mathbf{x}_i}$. Equation \eqref{refDensity2} holds true due to  \eqref{refDensity} and the fact that for
$$\mathbf{x}=\begin{pmatrix}
\mathbf{x}_1\\
\mathbf{x}_2
\end{pmatrix}\sim\mathcal{N}\left(
\begin{pmatrix}
\boldsymbol{\mu}_1\\
\boldsymbol{\mu}_2
\end{pmatrix}
,\begin{pmatrix}
\boldsymbol\Sigma_{11} & \boldsymbol\Sigma_{12}\\
\boldsymbol\Sigma_{21} & \boldsymbol\Sigma_{22}
\end{pmatrix}\right),$$
$(\mathbf{x}_1|\mathbf{x}_2=\mathbf{a})\sim\mathcal{N}(\bar{\boldsymbol{\mu}},\bar{\boldsymbol\Sigma})$ with $\bar{\boldsymbol{\mu}}=\boldsymbol{\mu}_1+\boldsymbol\Sigma_{12}\boldsymbol\Sigma_{22}^{-1}(\mathbf{a}-\boldsymbol{\mu}_2)$ and $\bar{\boldsymbol\Sigma}=\boldsymbol\Sigma_{11}-\boldsymbol\Sigma_{12}\boldsymbol\Sigma_{22}^{-1}\boldsymbol\Sigma_{21}$.

Following Datta et al. \cite{Datta2016}, the density $\tilde{p}(\mathbf{y}_{S})$ can be extended to a stochastic process that satisfies Komogorov's consistency criteria.
Let $\mathbf{u}\in X\setminus S$.

Further, let $N(\mathbf{u})$ be the set of $m$-nearest neighbors of $\mathbf{u}$ in $S$. For any finite set $U=\{\mathbf{u}_1,\ldots,\mathbf{u}_r\}$ with $S\cap U=\emptyset$ we define the nearest neighbor density of $\mathbf{y}_{U}$ conditional on $\mathbf{y}_{S}$ as
\begin{align}
\tilde{p}(\mathbf{y}_{U}|\mathbf{y}_{S})=\prod\limits_{i=1}^r p(y(\mathbf{u}_i)|\mathbf{y}_{N(\mathbf{u}_i)}). \label{refDensity3}
\end{align}
This conditional density is similar to \eqref{refDensity} except that all the neighbor sets are subsets of $S$. 
Indeed, \eqref{refDensity} and  \eqref{refDensity3} are sufficient to describe the joint nearest neighbor density of any finite set $V=\{\mathbf{v}_1,\ldots,\mathbf{v}_n\}\subseteq X$
\begin{align}
\tilde{p}(\mathbf{y}_{V})=\int\tilde{p}(\mathbf{y}_{U}|\mathbf{y}_{S})\tilde{p}(\mathbf{y}_{S})~\prod\limits_{\{\mathbf{s}_i\in S\setminus V\}}d\mathbf{y}(\mathbf{s}_i) ,\label{refDensity4}
\end{align}
where $U=V\setminus S$. 
If $U=\emptyset$, then  \eqref{refDensity3} implies that $\tilde{p}(\mathbf{y}_{U}|\mathbf{y}_{S})=1$ in  \eqref{refDensity4}. 
If $S\setminus V=\emptyset$ the integration in  \eqref{refDensity4} is not needed. 
Moreover, Datta et al. \cite{Datta2016} show that $\tilde{p}(\mathbf{y}_{V})$ is the density of a zero mean GP with covariance function,
\begin{align}
\tilde{C}(\mathbf{v}_1,\mathbf{v}_2)&=\begin{cases}
\tilde{C}_{\mathbf{x}_i,\mathbf{x}_j}, &\text{if } \mathbf{v}_1=\mathbf{x}_i,\mathbf{v}_2=\mathbf{x}_j \text{ are both in } S,\\
\mathbf{B}_{\mathbf{v}_1}\tilde{\mathbf{C}}_{N(\mathbf{v}_1),\mathbf{x}_j},  &\text{if } \mathbf{v}_1\notin S,\mathbf{v}_2=\mathbf{x}_j\in S,\\
\mathbf{B}_{\mathbf{v}_1}\tilde{\mathbf{C}}_{N(\mathbf{v}_1),N(\mathbf{v}_2)}\mathbf{B}_{\mathbf{v}_2}^T+\delta_{(\mathbf{v}_1=\mathbf{v}_2)}F_{\mathbf{v}_1}, &\text{if }  \mathbf{v}_1,\mathbf{v}_2\notin S,\end{cases}
\nonumber \\
\intertext{where}
\tilde{\mathbf{C}}_{S} &= \mathbf{B}_{S}^{-1}\mathbf{F}_S(\mathbf{B}_S^T)^{-1}, \label{eq:cstilde}
 \\
\intertext{with}
\mathbf{B}_{S} &= \begin{pmatrix}
1\\
&1&&\mathcal{O}\\
&&\ddots\\
&B_{\mathbf{x}_i,j}&&1\\
&&&&&1 
\end{pmatrix}, \qquad
B_{\mathbf{x}_i,j}=\begin{cases}
1, &  \text{if } i=j,\\
-\mathbf{B}_{\mathbf{x}_i}[l] ,& \text{if } \mathbf{x}_j = \mathbf{x}_i^l\text{ for some $l$},\\
0, & \text{else},
\end{cases} \nonumber
\\ 
\mathbf{B}_{\mathbf{v}_i}&=\mathbf{C}_{\mathbf{v}_i,N(\mathbf{v}_i)}\mathbf{C}_{N(\mathbf{v}_i)}^{-1},\nonumber
\\
\mathbf{F}_{S}&=\operatorname{diag}(F_{\mathbf{x}_1},\ldots,F_{\mathbf{x}_n}), \nonumber
\\
F_{\mathbf{v}_i}&=C_{\mathbf{v}_i}-\mathbf{B}_{\mathbf{v}_i}\mathbf{C}_{N(\mathbf{v}_i),\mathbf{v}_i}, \nonumber
\end{align}
where 
$\mathbf{x}_j = \mathbf{x}_i^l$ means that $\mathbf{x}_j$ is the $l$-th neighbor of $\mathbf{x}_i$ ($N(\mathbf{x}_i)=\left\{\mathbf{x}_i^1,\mathbf{x}_i^2,\ldots,\mathbf{x}_i^l,\ldots,\mathbf{x}_i^{|N(\mathbf{x}_i)|}\right\}$) , $\mathbf{v}_1,\mathbf{v}_2\in X$, $\tilde{\mathbf{C}}_{A,B}$ denotes submatrices of $\tilde{\mathbf{C}}$ indexed by the elements of the sets $A$ and $B$.
The Kronecker delta is denoted by $\delta_{(\mathbf{v}_1=\mathbf{v}_2)}$.

\subsection{Reference Prior}
\label{ss:refPrior}

Paulo \cite{paulo2005} computes the general form of the reference prior for a broad class of Gaussian processes. 
Assume that $y(\cdot)$ is a GP on $X\subseteq\mathbb{R}^d$ with mean and covariance given by
\begin{align}
\mu(\mathbf{x})=\mathbf{x}^T\boldsymbol\beta, \qquad C(\mathbf{x}_1,\mathbf{x}_2)=\sigma^2K(\mathbf{x}_1,\mathbf{x}_2|\boldsymbol{\xi}),\nonumber
\end{align}
where $\boldsymbol\beta\in\mathbb{R}^d$, $\boldsymbol\xi\in\mathbb{R}_+^r$ stands for a vector of additional correlation parameters, and $\sigma^2>0$ is the process variance. 
Some additional notation is required to define the likelihood function corresponding to the observed data $D$:
\begin{align}
\mathbf{X}&=(\mathbf{x}_1 \cdots \mathbf{x}_n)^T,\nonumber\\
\boldsymbol\eta&=(\sigma^2,\boldsymbol{\beta}^T,\boldsymbol{\xi}^T)^ T,\nonumber\\
\mathbf K_{S}&=[K(\mathbf{x}_i,\mathbf{x}_j|\boldsymbol{\xi})]_{\mathbf{x}_i,\mathbf{x}_j\in S}.\nonumber
\end{align}
Then, $\mathbf{y}_S|\boldsymbol\eta \sim \mathcal{N}(\mathbf{X}\boldsymbol\beta,\sigma^2\mathbf{ K}_{S})$, which implies that the likelihood is given by:
\begin{align}
L(\boldsymbol\eta|\mathbf{y}_S)\propto (\sigma^2)^{-\frac{n}{2}}|\mathbf{K}_{S}|^{-\frac{1}{2}}\operatorname{exp}\left\{-\frac{1}{2\sigma^2}(\mathbf{y}_S-\mathbf{X}\boldsymbol\beta)^T\mathbf{K}_{S}^{-1}(\mathbf{y}_S-\mathbf{X}\boldsymbol\beta)\right\} \nonumber.
\end{align}
When defining the reference prior, $(\sigma^2,\boldsymbol\xi^T)^T$ is considered as a parameter of interest, and $\boldsymbol\beta$ is considered as a  nuisance parameter. Moreover, the distribution of the reference prior $\pi^{R}(\boldsymbol\eta)$ is factored as $\pi^{R}(\boldsymbol\eta)=\pi^{R}(\boldsymbol\beta|\sigma^2,\boldsymbol\xi)\pi^{R}(\sigma^2,\boldsymbol\xi)$. 
The conditional prior of $\boldsymbol\beta$ is specified proportional to $1$, i.e., $\pi^{R}(\boldsymbol\beta|\sigma^2,\boldsymbol\xi)\propto 1$, because this is the Jeffreys-rule prior (see \cite{10.1214/09-STS284}), which is a noninformative prior, for the model at hand when $\sigma^2$ and $\boldsymbol\xi$ are considered to be known. 

Next, $\pi^{R}(\sigma^2,\boldsymbol\xi)$ is calculated as the Jeffreys-rule prior for the marginal experiment defined via the integrated likelihood function with respect to $\pi^{R}(\boldsymbol\beta|\sigma^2,\boldsymbol\xi)$. 
The integrated likelihood function is given by:
\begin{align}
L^{I}(\sigma^2,\boldsymbol\xi|\mathbf{y}_S)&=\int\limits_{\mathbb{R}^d}L(\boldsymbol\eta|\mathbf{y}_S)\pi^{R}(\boldsymbol\beta|\sigma^2,\boldsymbol\xi)~d\boldsymbol\beta\nonumber,\\
&=\int\limits_{\mathbb{R}^d}L(\boldsymbol\eta|\mathbf{y}_S)~d\boldsymbol\beta\nonumber,\\
&\propto (\sigma^2)^{-\frac{n-d}{2}}|\mathbf{K}_{S}|^{-\frac{1}{2}}||\mathbf{X}^T\mathbf{K}_{S}^{-1}\mathbf{X}|^{-\frac{1}{2}}\operatorname{exp}\left\{-\frac{S_{\boldsymbol\xi}^2}{2\sigma^2}\right\},\nonumber
\end{align}
where  $S_{\boldsymbol\xi}^2=\mathbf{y}_S^T\mathbf{Q}\mathbf{y}_S$, $\mathbf{Q}=\mathbf{K}_{S}^{-1}\mathbf{P}$, and $\mathbf{P}=\mathbf{I}-\mathbf{X}(\mathbf{X}^T\mathbf{K}_{S}^{-1}\mathbf{X})^{-1}\mathbf{X}^T\mathbf{K}_{S}^{-1}$.

Finally, the reference prior $\pi^{R}(\boldsymbol\eta)$, see also \citep{paulo2005}, is of the form $\pi^{R}(\boldsymbol\eta)\propto\frac{\pi^{R}(\boldsymbol\xi)}{\sigma^2}$ with $\pi^{R}(\boldsymbol\xi)\propto |\tilde{I}_{R}(\boldsymbol\xi)|^{\frac{1}{2}}$ where
\begin{align}
\mathbf{W}_{k}&=\frac{\partial\mathbf{K}_{S}}{\partial \xi_k}\mathbf{Q}, \qquad k=1,\ldots,r,  \nonumber\\
\tilde{I}_{R}(\boldsymbol\xi)&=
\begin{pmatrix}
n-d & \operatorname{tr}(\mathbf{W}_1)& \operatorname{tr}(\mathbf{W}_2) & \cdots & \operatorname{tr}(\mathbf{W}_r) \\
\operatorname{tr}(\mathbf{W}_1) & \operatorname{tr}(\mathbf{W}_1^2) &\operatorname{tr}(\mathbf{W}_1\mathbf{W}_{2}) & \cdots & \operatorname{tr}(\mathbf{W}_1\mathbf{W}_r) \\
\vdots &\vdots & \vdots & &\vdots\\
\operatorname{tr}(\mathbf{W}_{r}) & \operatorname{tr}(\mathbf{W}_1\mathbf{W}_{r})&\operatorname{tr}(\mathbf{W}_2\mathbf{W}_{r})&\cdots & \operatorname{tr}(\mathbf{W}_r^2)  \nonumber
\end{pmatrix}.
\end{align}

\subsection{Variable Selection Model}
\label{ss:vsm}
In this section, we formally introduce our variable selection model.
At first, we define the likelihood function. 
Afterward, in a Bayesian manner,  prior distributions are assigned to the model parameters.

\subsubsection{Likelihood Function}

Let $y\in\mathbb{R}$ denote a target variable of interest and let $\mathbf{x}\in\mathbb{R}^d$ denote covariates.
It is assumed that the relationship.
\begin{align}
y = \mathbf{x}^T\boldsymbol\beta + f(\mathbf{x})+\varepsilon,  
 \label{model}
\end{align}
holds, where $f$ denotes an unknown function and $\varepsilon$ additive noise. Moreover, let $\mathcal{A}$ denote a (random) subset of the index set $\left\{1,\ldots,d\right\}$, with $|\mathcal{A}|=k>0$. 
This set is used to specify the covariates that contribute to the regression model  \eqref{model}. For given $\mathcal{A}$ we assign a GP prior with zero mean and covariance function $\tilde{C}$ to the sum $z:= f+\varepsilon$:
\begin{align}
z\sim GP(0,\tilde{C}(\cdot,\cdot|\sigma^2,\gamma,\rho,\mathcal{A})), 
 \label{GPprior}
\end{align}
with $\sigma^2> 0,~ \rho > 0$ and $\gamma\in(0,1)$.
In particular, the GP prior $GP(0,\tilde{C}(\cdot,\cdot|\sigma^2,\gamma,\rho, \mathcal{A}))$  \eqref{GPprior} is defined as the NNGP derived from the parent $GP(0,C(\cdot,\cdot|\sigma^2,\gamma,\rho, \mathcal{A}))$, with
\begin{align}
C(\mathbf{x},\mathbf{x}^\prime|\sigma^2,\gamma,\rho, \mathcal{A})&=\sigma^2\underbrace{\left[\delta_{(\mathbf{x}=\mathbf{x}^\prime)}(1-\gamma)+\gamma K(\mathbf{x},\mathbf{x}^\prime|\rho, \mathcal{A})\right]}_{=\mathcal{K}(\mathbf{x},\mathbf{x}^\prime|\gamma,\rho, \mathcal{A})}, \label{covParent} \\
\intertext{where}
K(\mathbf{x},\mathbf{x}^\prime|\rho, \mathcal{A}) &= \left(1+\frac{\sqrt{5}d_{\mathcal{A}}}{\rho}+\frac{5d_{\mathcal{A}}^2}{3\rho^2}\right)\operatorname{exp}\left(-\frac{\sqrt{5}d_{\mathcal{A}}}{\rho}\right), \nonumber \\
\intertext{with}
d_{\mathcal{A}} &= \sqrt{\sum\limits_{i\in\mathcal{A}}(x_i-x_i^\prime)^{2}}. \label{eq:euclideanDistance}
\end{align}
Therefore, the parent GP has a Mat{\'e}rn-$\frac{5}{2}$ covariance function (\cite{matern1966spatial}).
The Mat{\'e}rn covariance function can be seen as a generalization of the Gaussian radial basis function.

The Euclidean distance function  $d_{\mathcal{A}}$ \eqref{eq:euclideanDistance}  considers only predictors with indices in $\mathcal{A}$.
Function $d_{\mathcal{A}}$ is also used to determine the neighbor sets required for the computation of the NNGP.

It is noteworthy that the parent GP results from assigning a zero mean GP with a covariance function $\sigma_{f}^2K(\mathbf{x},\mathbf{x}^\prime|\rho,\mathcal{A})$ to $f$ while  assuming the noise $\varepsilon$ to be $iid$ $\mathcal{N}(0,\sigma_{\varepsilon}^2)$. The reparameterization $\sigma^2=\sigma_{f}^2+\sigma_{\varepsilon}^2, \gamma=1-\frac{\sigma_{\varepsilon}^2}{\sigma^2}$ in (\ref{covParent}) gives the following:
\begin{align}
C(\mathbf{x},\mathbf{x}^\prime|\sigma_{f}^2,\sigma_{\varepsilon}^2,\rho,\mathcal{A})&=\sigma_{\varepsilon}^2\delta_{(\mathbf{x}=\mathbf{x}^\prime)}+\sigma_{f}^2K(\mathbf{x},\mathbf{x}^\prime|\rho, \mathcal{A}) \nonumber,\\
&=\sigma^2\left[\frac{\sigma_{\varepsilon}^2}{\sigma^2}\delta_{(\mathbf{x}=\mathbf{x}^\prime)}+\frac{\sigma^2-\sigma_{\varepsilon}^2}{\sigma^2}K(\mathbf{x},\mathbf{x}^\prime|\rho,\mathcal{A})\right],\nonumber\\
&=\sigma^2\left[(1-\gamma)\delta_{(\mathbf{x}=\mathbf{x}^\prime)}+\gamma K(\mathbf{x},\mathbf{x}^\prime|\rho,\mathcal{A})\right]. \nonumber
\end{align}

Above specifications directly imply that $y$ follows a GP with mean function $\mu(\mathbf{x})=\mathbf{x}^T\boldsymbol\beta$ and covariance function $\tilde{C}(\mathbf{x},\mathbf{x}^{\prime}|\sigma^2,\gamma,\rho,\mathcal{A})$. Let $D=\{(\mathbf{x}_i,y_i)|i=1\ldots,n\}$ denote some training data and, further, let $S = \{\mathbf{x}_1,\ldots,\mathbf{x}_n\}$ denote the corresponding set of covariate constellations (experimental design). Consider the following additional notations:
\begin{align}
\tilde{\mathbf C}_{S}&=[\tilde{C}(\mathbf{x}_i,\mathbf{x}_j|\sigma^2,\gamma,\rho,\mathcal{A})]_{\mathbf{x}_i,\mathbf{x}_j\in S}. \nonumber
\intertext{We recall from \eqref {eq:cstilde}  that $\tilde{\mathbf C}_{S}$ is defined as:}
\tilde{\mathbf{C}}_{S}& = \mathbf{B}_{S}^{-1}\mathbf{F}_S(\mathbf{B}_S^T)^{-1} , \nonumber
\intertext{with}
\mathbf{B}_{S} &= \begin{pmatrix}
1\\
&1&&\mathcal{O}\\
&&\ddots\\
&B_{\mathbf{x}_i,j}&&1\\
&&&&&1 
\end{pmatrix},\qquad
B_{\mathbf{x}_i,j}=\begin{cases}
1,& \text{if } i=j,\\
-\mathbf{B}_{\mathbf{x}_i}[l], & \text{if }\mathbf{x}_j = \mathbf{x}_i^l\text{ for some $l$},\\
0, & \text{else},
\end{cases} \nonumber\\
\mathbf{B}_{\mathbf{x}_i}&=\mathbf{C}_{\mathbf{x}_i,N(\mathbf{x}_i)}\mathbf{C}_{N(\mathbf{x}_i)}^{-1}=\sigma^2\boldsymbol{\mathcal{K}}_{\mathbf{x}_i,N(\mathbf{x}_i)}(\sigma^2)^{-1}\boldsymbol{\mathcal{K}}_{N(\mathbf{x}_i)}^{-1},  \nonumber\\
&={\boldsymbol{\mathcal{K}}_{\mathbf{x}_i,N(\mathbf{x}_i)}\boldsymbol{\mathcal{K}}_{N(\mathbf{x}_i)}^{-1}  }~\quad\text{(independent of $\sigma^2$)}, \nonumber\\
\mathbf{F}_{S}&=\operatorname{diag}(F_{\mathbf{x}_1},\ldots,F_{\mathbf{x}_n}), \nonumber\\
F_{\mathbf{x}_i}&=C_{\mathbf{x}_i}-\mathbf{B}_{\mathbf{x}_i}\mathbf{C}_{N(\mathbf{x}_i),\mathbf{x}_i} ,\nonumber\\
&=\sigma^2\mathcal{K}_{\mathbf{x}_i}-\mathbf{B}_{\mathbf{x}_i}\sigma^2\boldsymbol{\mathcal{K}}_{N(\mathbf{x}_i),\mathbf{x}_i} ,\nonumber\\
&= \sigma^2\underbrace{\left[\mathcal{K}_{\mathbf{x}_i}-\mathbf{B}_{\mathbf{x}_i}\boldsymbol{\mathcal{K}}_{N(\mathbf{x}_i),\mathbf{x}_i}\right]}_{=\mathcal{F}_{\mathbf{x}_i}}^{\text{(independent of $\sigma^2$)}}. \nonumber
\intertext{Finally, we find that:}
\tilde{\mathbf{C}}_{S} &= \sigma^2\underbrace{\mathbf{B}_{S}^{-1}\operatorname{diag}(\mathcal{F}_{\mathbf{x}_1},\ldots,\mathcal{F}_{\mathbf{x}_n})(\mathbf{B}_{S}^T)^{-1}}_{=:\boldsymbol{\tilde{\mathcal{K}}}_{S}}, \nonumber\\
&= \sigma^2 \boldsymbol{\tilde{\mathcal{K}}}_{S}.\nonumber
\end{align}
Note that $\boldsymbol{\tilde{\mathcal{K}}}_{S}$ is independent on $\sigma^2$ while it is  dependent of $\rho$, $\gamma$, and $\mathcal{A}$.
Thus, $\mathbf{y}_S|\sigma^2,\gamma,\rho, \mathcal{A} \sim \mathcal{N}(\mathbf{X}\boldsymbol\beta,\sigma^2\boldsymbol{\tilde{\mathcal{K}}}_{S})$, which implies that the likelihood function is given by:
\begin{align}
L(\boldsymbol\beta,\sigma^2,\gamma,\rho, \mathcal{A}|\mathbf{y}_S)\propto (\sigma^2)^{-\frac{n}{2}}|\boldsymbol{\tilde{\mathcal{K}}}_{S}|^{-\frac{1}{2}}\operatorname{exp}\left\{-\frac{1}{2\sigma^2}(\mathbf{y}_S-\mathbf{X}\boldsymbol\beta)^T\boldsymbol{\tilde{\mathcal{K}}}_{S}^{-1}(\mathbf{y}_S-\mathbf{X}\boldsymbol\beta)\right\}. \nonumber
\end{align}

\subsubsection{Reference Prior}
For given $\mathcal{A}$ we denote the components of $\boldsymbol\beta$ with indices in $\mathcal{A}$ as $\beta_{\mathcal{A}}$ and the complementary ones 
 as $\boldsymbol\beta_{\mathcal{A}^C}$. 
 We set $\boldsymbol\beta_{\mathcal{A}^C}$  equal to zero and assign a uniform prior to $\beta_{\mathcal{A}}$.
 The joint prior density with respect to the measure $(\lambda+\delta_0)^d$ is given by:
\begin{align}
p(\boldsymbol\beta|\mathcal{A})=p(\boldsymbol\beta_{\mathcal{A}}|\mathcal{A})p(\boldsymbol\beta_{\mathcal{A}^C}|\mathcal{A})\propto I_{\{\boldsymbol\beta_{\mathcal{A}^C}=\mathbf{0}\}}(\boldsymbol\beta_{\mathcal{A}^C})I_{\{\beta_{i}\neq 0|\forall i \in \mathcal{A}\}}(\boldsymbol\beta_{\mathcal{A}}), \nonumber
\end{align}
where $\lambda$ denotes the Lebesgue measure and $\delta_0$ denotes the Dirac measure concentrated at zero. Thus, the  integrated likelihood (compare to Section \ref{ss:refPrior}) is given by:
\begin{align}
L^{I}(\sigma^2,\rho,\gamma|\mathbf{y}_S, \mathcal{A})&=\int\limits_{\mathbb{R}^d}L(\sigma^2,\rho,\gamma,\boldsymbol\beta|\mathbf{y}_S, \mathcal{A})p(\boldsymbol\beta|\mathcal{A})~d\lambda(\boldsymbol\beta_\mathcal{A})d(\delta_0(\boldsymbol\beta_\mathcal{A}^C)), \nonumber\\
&\propto (\sigma^2)^{-\frac{n-|\mathcal{A}|}{2}}|\boldsymbol{\tilde{\mathcal{K}}}_{S}|^{-\frac{1}{2}}||\mathbf{X}_{\mathcal{A}}^T\boldsymbol{\tilde{\mathcal{K}}}_{S}^{-1}\mathbf{X}_{\mathcal{A}}|^{-\frac{1}{2}}\operatorname{exp}\left\{-\frac{S_{\boldsymbol\lambda}^2}{2\sigma^2}\right\} ,\nonumber
\end{align}
where $\boldsymbol\lambda = (\rho,\gamma,\mathcal{A})^T$, $S_{\boldsymbol\lambda}^2=\mathbf{y}_S^T\mathbf{Q}\mathbf{y}_S$, $\mathbf{Q}=\boldsymbol{\tilde{\mathcal{K}}}_{S}^{-1}\mathbf{P}$, $\mathbf{P}=\mathbf{I}-\mathbf{X}_{\mathcal{A}}(\mathbf{X}_{\mathcal{A}}^T\boldsymbol{\tilde{\mathcal{K}}}_{S}^{-1}\mathbf{X}_{\mathcal{A}})^{-1}\mathbf{X}_{\mathcal{A}}^T\boldsymbol{\tilde{\mathcal{K}}}_{S}^{-1}$, and $\mathbf{X}_{\mathcal{A}}$ denotes the corresponding to the set $\mathcal{A}$ reduced design matrix.

In accordance with Section \ref{ss:refPrior} we use the prior  
\begin{align}
p(\sigma^2, \rho, \gamma) &\propto \frac{|\tilde{I}_{R}(\rho, \gamma)|^{\frac{1}{2}}}{\sigma^2}, \nonumber
\intertext{where}
\mathbf{W}_{\rho}&=\frac{\partial\boldsymbol{\tilde{\mathcal{K}}}_{S}}{\partial \xi_k}\mathbf{Q} ,\label{deriv1}\\
\mathbf{W}_{\gamma}&=\frac{\partial\boldsymbol{\tilde{\mathcal{K}}}_{S}}{\partial \gamma}\mathbf{Q} ,\label{deriv2}\\
\tilde{I}_{R}(\rho, \gamma)&=
\begin{pmatrix}
n-|\mathcal{A}| & \operatorname{tr}(\mathbf{W}_{\rho}) & \operatorname{tr}(\mathbf{W}_{\gamma}))\\
\operatorname{tr}(\mathbf{W}_{\rho}) & \operatorname{tr}(\mathbf{W}_{\rho}^2)  & \operatorname{tr}(\mathbf{W}_{\rho}\mathbf{W}_{\gamma})\\
\operatorname{tr}(\mathbf{W}_{\gamma}) & \operatorname{tr}(\mathbf{W}_{\rho}\mathbf{W}_{\gamma})& \operatorname{tr}(\mathbf{W}_{\rho}^2) 
\end{pmatrix}\nonumber.
\end{align}

Details concerning the computation of the derivatives presented in Equation \eqref{deriv1} and Equation \eqref{deriv2} are provided in Appendix \ref{secA1}.

\subsubsection{Prior for the Random Set}
\label{ss:priorA}
As in Posch et al. \cite{POSCH2020}, we propose the following prior for $\mathcal{A}$:
\begin{align}
   & p(\mathcal{A}=\{\alpha_1,\ldots,\alpha_{k}\})\propto (p_{\alpha_1}+\ldots+p_{\alpha_k})\frac{1}{k}\tilde{p}(k), \nonumber \\
   \intertext{where}
&\{p_{\alpha_1},\ldots,p_{\alpha_k}\}\subseteq\{p_1,\ldots,p_d\} \text{ with } \sum_{i=1}^{d}p_i=1 \text{ and } p_i\geq 0  \text{ for } i=1,\ldots,d  \text{ and,} \nonumber \\
&\tilde{p}:\{1,\ldots,d\}\rightarrow\mathbb{R}_{0}^{+}. \nonumber 
\end{align}
The mapping $\tilde{p}$ can be arbitrarily chosen and should represent the a priori belief in the number of predictors contributing to the model. Further,  the parameters $p_1,\ldots,p_d$ are used to represent the a priori belief in the importance of the predictors $x_1,\ldots,x_d$ to the model.

\subsection{Model Inference}
\label{ss:inference}
Clearly, the joint posterior  density
 $p(\boldsymbol\beta,\sigma^2,\gamma,\rho,\mathcal{A}|\mathbf{y}_S,\mathbf{X})$ is analytically intractable. To overcome this limitation, we propose a Metropolis-within-Gibbs algorithm to sample from this distribution. 
 The algorithm produces a Markov Chain with the true posterior as the limiting distribution. In particular, the algorithm includes three main steps:
\begin{itemize}
\item \textbf{Step 1:} Sample from the conditional posterior $p(\boldsymbol\beta_{t+1},\mathcal{A}_{t+1}|\mathbf{y}_S,\mathbf{X},\sigma_{t}^2,\gamma_{t},\rho_{t})$ using a classical Metropolis-Hastings (MH) algorithm.
\item \textbf{Step 2:} Sample from the conditional posterior $p(\sigma_{t+1}^2|\mathbf{y}_S,\mathbf{X},\gamma_{t},\rho_{t},\boldsymbol\beta_{t+1},\mathcal{A}_{t+1})$.
\item \textbf{Step 3:} Sample from the conditional posterior $p(\gamma_{t+1},\rho_{t+1}|\mathbf{y}_S,\mathbf{X},\boldsymbol\beta_{t+1},\mathcal{A}_{t+1},\sigma_{t+1}^2)$ using Hamiltonian MCMC (HMC).
\end{itemize}
In order to run the procedure one has to define an initial sample $\boldsymbol{\beta}_0,\sigma_0^2,\gamma_0,\rho_0,\mathcal{A}_0$ that lies in the support of $p(\boldsymbol\beta,\sigma^2,\gamma,\rho,\mathcal{A}|\mathbf{y}_S,\mathbf{X})$. 
Additionally, some tuning parameters for the proposal distribution of $\mathcal{A}$ are required, and also the step size $S$ and the number of steps $L$ for the leapfrog method, that is used within \textbf{Step 3} (HMC).
Details concerning the three steps are provided below.

\subsubsection{Step 1:}
To specify the proposal for the random set $\mathcal{A}$ at first a Bernoulli distributed random variable $c_{h}$ is introduced (as suggested by \cite{POSCH2020})
\begin{equation}
c_{h}\sim \text{Bernoulli}(p_h), \label{propA1}
\end{equation}
where $p_{h}\in [0,1]$ can be considered as a tuning parameter. The event $c_h=1$ means that the model size (number of predictors used) changes, i.e., $k_{t+1}\neq k_t$.
Moreover, a random variable $\alpha$ with support on the index set $\{1,\ldots,d\}$ is introduced in order to describe the model transition probabilities in case of changing model size. 
A realization of the conditional random variable $\alpha|\mathcal{A}_t$ corresponds to the index of a predictor that is going to be added to or removed from the model. 
Transitions between models, which differ by two or more parameters, are not allowed. The probability mass function of $\alpha|\mathcal{A}_t$ is defined as
\begin{equation}
q(\alpha|\mathcal{A}_t)=\text{I}_{\{1,\ldots,d\}}(\alpha)\begin{cases}
\tilde{p}_{\alpha}, &\text{if } k_t>1,\alpha\notin\mathcal{A}_t,\\
\frac{\sum\limits_{i\in\mathcal{A}_t}\tilde{p}_i}{\tilde{p}_\alpha\sum\limits_{i\in\mathcal{A}_t}1/\tilde{p}_i},&\text{if } k_t>1,\alpha\in\mathcal{A}_t,\\
0,&\text{if } k_t=1,\alpha\in\mathcal{A}_t,\\
\frac{\tilde{p}_{\alpha}}{\sum\limits_{i\in\{1,\ldots,p\}\setminus\mathcal{A}_t}\tilde{p}_i}, &\text{else},
\end{cases}\label{propA2}
\end{equation}
where the parameters $\tilde{p}_1,\ldots,\tilde{p}_d$ are greater or equal to zero, sum up to one, and can but must not be identical to the parameters $p_1,\ldots,p_d$ already used in the prior $p(\mathcal{A})$. Using \eqref{propA1} and  \eqref{propA2} the proposal distribution $q(\mathcal{A}_{t+1}|\mathcal{A}_{t})$ is finally defined by:
\begin{align}
&q(\mathcal{A}_{t+1}|\mathcal{A}_{t},c_h=1)=q([\mathcal{A}_{t+1}\setminus\mathcal{A}_{t}]\cup[\mathcal{A}_{t}\setminus\mathcal{A}_{t+1}]|\mathcal{A}_{t})\text{I}_{\{1\}}(|k_{t+1}-k_t|),  \nonumber\\
&q(\mathcal{A}_{t+1}|\mathcal{A}_{t},c_h=0)=\begin{cases}
1, &\text{if } \mathcal{A}_{t+1}=\mathcal{A}_t,\\
0, &\text{else},
\end{cases}  \nonumber\\
&q(\mathcal{A}_{t+1}|\mathcal{A}_{t})=\sum\limits_{c_h\in\{0,1\}}q(\mathcal{A}_{t+1}, c_h|\mathcal{A}_{t}),  \nonumber\\
&\hspace{1.7cm}= \sum\limits_{c_h\in\{0,1\}}q(\mathcal{A}_{t+1}|c_h, \mathcal{A}_{t})q(c_h). \nonumber
\end{align}
By straightforward algebraic manipulation, it's evident that:
\begin{align}
\boldsymbol\beta_{t+1,\mathcal{A}_{t+1}}|\mathbf{y}_S,\mathbf{X},\sigma_{t}^2,\gamma_{t},\rho_{t},\mathcal{A}_{t+1}&\sim\mathcal{N}(\boldsymbol\mu =\mathbf{\Sigma}\mathbf{X}_{\mathcal{A}_{t+1}}^T\mathbf{\tilde{C}}_S^{-1}\mathbf{y}_S, \mathbf{\Sigma}=(\mathbf{X}_{\mathcal{A}_{t+1}}^T\mathbf{\tilde{C}}_S^{-1}\mathbf{X}_{\mathcal{A}_{t+1}})^{-1}),\label{propBeta1}\\
\boldsymbol\beta_{t+1,\mathcal{A}_{t+1}^C}|\mathbf{y}_S,\mathbf{X},\sigma_{t}^2,\gamma_{t},\rho_{t},\mathcal{A}_{t+1}&\sim\delta_0. \label{propBeta2}
\end{align}
We use the conditional posterior \eqref{propBeta1} and \eqref{propBeta2} as proposal for $\boldsymbol\beta_{t+1}$. The Hastings ratio $r$ is then given by \eqref{eq:hastingRatio}.
\begin{align}
r = \frac{L(\boldsymbol\beta_{t+1},\sigma_{t}^2,\gamma_{t},\rho_{t}, \mathcal{A}_{t+1}|\mathbf{y}_S)p(\boldsymbol\beta_{t+1}|\mathcal{A}_{t+1})p(\mathcal{A}_{t+1})q(\boldsymbol\beta_{t}|\sigma_{t}^2,\gamma_{t},\rho_{t},\mathcal{A}_{t})q(\mathcal{A}_{t}|\mathcal{A}_{t+1})}{L(\boldsymbol\beta_{t},\sigma_{t}^2,\gamma_{t},\rho_{t}, \mathcal{A}_{t}|\mathbf{y}_S)p(\boldsymbol\beta_{t}|\mathcal{A}_{t})p(\mathcal{A}_{t})q(\boldsymbol\beta_{t+1}|\sigma_{t}^2,\gamma_{t},\rho_{t},\mathcal{A}_{t+1})q(\mathcal{A}_{t+1}|\mathcal{A}_{t})}. \label{eq:hastingRatio}
\end{align}

\subsubsection{Step 2:}
The conditional posterior $p(\sigma_{t+1}^2|\mathbf{y}_S,\mathbf{X},\gamma_{t},\rho_{t},\boldsymbol\beta_{t+1},\mathcal{A}_{t+1})$ can be derived in a closed analytical form. 
Upon straightforward algebraic manipulation, it comes out as an inverse gamma (IG) distribution:
\begin{align}
\sigma_{t+1}^2|\mathbf{y}_S,X,\gamma_{t},\rho_{t},\boldsymbol\beta_{t+1},\mathcal{A}_{t+1}\sim IG(\alpha=\frac{n}{2},\beta =\frac{1}{2}(\mathbf{y}_S-\mathbf{X}\boldsymbol\beta)^T\boldsymbol{\tilde{\mathcal{K}}}_S^{-1}(\mathbf{y}_S-\mathbf{X}\boldsymbol\beta)).
\end{align}

\subsubsection{Step 3:}
For the HMC algorithm we reparameterize the posterior $p(\gamma_{t+1},\rho_{t+1}|\mathbf{y}_S,\mathbf{X},\boldsymbol\beta_{t+1},\mathcal{A}_{t+1},\sigma_{t+1}^2)$, such that the support of the reparameterized distribution is unbounded. This can be easily done using Jacobi transformation:
\begin{align}
\gamma_{t+1} &= \frac{1}{1+\operatorname{exp}(-\tilde{\gamma}_{t+1})} \text{ with } \tilde{\gamma}_{t+1}\in\mathbb{R},\nonumber\\
\rho_{t+1}&=\operatorname{log}(1+\operatorname{exp}(\tilde{\rho}_{t+1})) \text{ with } \tilde{\rho}_{t+1}\in\mathbb{R},\nonumber\\
p(\tilde{\gamma}_{t+1},\tilde{\rho}_{t+1}|\mathbf{y}_S,\mathbf{X},\boldsymbol\beta_{t+1},\mathcal{A}_{t+1},\sigma_{t+1}^2)&\propto L(\boldsymbol\beta_{t+1},\sigma_{t+1}^2,\gamma_{t+1},\rho_{t+1}, \mathcal{A}_{t+1}|\mathbf{y}_S)p(\rho_{t+1},\gamma_{t+1})\nonumber\\
&\frac{\operatorname{exp}(\tilde{\rho}_{t+1}-\tilde{\gamma}_{t+1})}{(1+\operatorname{exp}(-\tilde{\gamma}_{t+1})^2)(1+\operatorname{exp}(\tilde{\rho}_{t+1}))}.\label{potEnergy}
\end{align}
Next we specify the so-called potential energy function:
\begin{align}
E(\tilde{\rho}_{t+1},\tilde{\gamma}_{t+1})&=-\operatorname log(p(\tilde{\gamma}_{t+1},\tilde{\rho}_{t+1}|\mathbf{y}_S,\mathbf{X},\boldsymbol\beta_{t+1},\mathcal{A}_{t+1},\sigma_{t+1}^2)) \nonumber
\end{align}
Due to Equation \eqref{potEnergy} $E(\cdot|\cdot)$ is known up to an additive constant that vanishes if derivatives are computed. Moreover, the kinetic energy is defined as the negative logarithm of the density of a zero mean normal distribution with a diagonal covariance matrix:
\begin{align}
K(\boldsymbol{v}=(v_{\tilde{\rho}},v_{\tilde{\gamma}})^T)=c+\frac{1}{2}\left(\frac{v_{\tilde{\rho}}^2}{m_{\tilde{\rho}}}+\frac{v_{\tilde{\gamma}}^2}{m_{\tilde{\gamma}}}\right).  \nonumber
\end{align}
The parameters $m_{\tilde{\rho}}$ and $m_{\tilde{\gamma}}$ are tuning parameters of the algorithm and, further, $c$ denotes an additive constant. Consequently, the partial derivatives of the kinetic energy function are given by:
\begin{align}
\frac{\partial K(\boldsymbol{v})}{\partial v_{\tilde{\rho}}} = \frac{v_{\tilde{\rho}}}{m_{\tilde{\rho}}},\qquad\frac{\partial K(\boldsymbol{v})}{\partial v_{\tilde{\gamma}}} = \frac{v_{\tilde{\gamma}}}{m_{\tilde{\gamma}}}.  \nonumber
\end{align}
With above considerations the HMC algorithm proposes a sample from the conditional posterior by applying the following procedure:
\begin{enumerate}
\item Draw a sample $\boldsymbol v=(v_{\tilde{\rho}},v_{\tilde{\gamma}})^T$ from $\mathcal{N}(\mathbf{0},\operatorname{diag}(m_{\tilde{\rho}},m_{\tilde{\gamma}}))$.
\item Use the leapfrog method to simulate $\tilde{\rho}^*,\tilde{\gamma}^*,v_{\tilde{\rho}}^*,v_{\tilde{\gamma}}^*$.
\begin{enumerate}[(i)]
\item Set $\boldsymbol v^*=\boldsymbol v, \tilde{\rho}^*=\tilde{\rho}_t,\tilde{\gamma}^*=\tilde{\gamma}_t$.
\item For $j=1,\ldots,L:$
\begin{enumerate}[(a)]
\item $v_{\tilde{\rho}}^*=v_{\tilde{\rho}}^*-\frac{\varepsilon}{2}\frac{\partial E}{\partial\tilde{\rho}^*}(\tilde{\rho}^*,\tilde{\gamma}^*),~v_{\tilde{\gamma}}^*=v_{\tilde{\gamma}}^*-\frac{\varepsilon}{2}\frac{\partial E}{\partial\tilde{\gamma}^*}(\tilde{\rho}^*,\tilde{\gamma}^*)$.
\item $\tilde{\rho}^* = \tilde{\rho}^*+\varepsilon \frac{v_{\tilde{\rho}}^*}{m_{\tilde{\rho}}},\tilde{\gamma}^* = \tilde{\gamma}^*+\varepsilon \frac{v_{\tilde{\gamma}}^*}{m_{\tilde{\gamma}}}$.
\item $v_{\tilde{\rho}}^*=v_{\tilde{\rho}}^*-\frac{\varepsilon}{2}\frac{\partial E}{\partial\tilde{\rho}^*}(\tilde{\rho}^*,\tilde{\gamma}^*),~v_{\tilde{\gamma}}^*=v_{\tilde{\gamma}}^*-\frac{\varepsilon}{2}\frac{\partial E}{\partial\tilde{\gamma}^*}(\tilde{\rho}^*,\tilde{\gamma}^*)$.
\end{enumerate}
\end{enumerate} 
\item Calculate the Hastings ratio $r = \operatorname{exp}[K(\boldsymbol v)+E(\tilde{\rho}_{t},\tilde{\gamma}_{t})-K(\boldsymbol v^*)-E(\tilde{\rho}^*,\tilde{\gamma}^*)]$.
\item With probability $\operatorname{min}(r,1)$ set $\tilde{\rho}_{t+1}=\tilde{\rho}^*,\tilde{\gamma}_{t+1}=\tilde{\gamma}^*$. Otherwise set $\tilde{\rho}_{t+1}=\tilde{\rho}_{t},\tilde{\gamma}_{t+1}=\tilde{\gamma}_{t}$.
\end{enumerate}

\subsection{Prediction}
\label{ss:prediction}

Let $\mathbf{X}^*=(\mathbf{x}_1^*\cdots \mathbf{x}_v^*)^T$ be the design matrix corresponding to some predictor specifications for which predictions $\widehat{\mathbf{y}}^*$ are required. Note that $\mathbf{y}_S$ and $\mathbf{y}^*$ are joint normally distributed, i.e.,
\begin{align}
(\mathbf{y}^*,
\mathbf{y}_S)^T
|\mathbf{X}^*,\mathbf{X},\boldsymbol\beta, \sigma^2, \gamma, \rho, \mathcal{A}
\sim
\mathcal{N}\begin{pmatrix}
\begin{pmatrix}
\mathbf{X}^*\boldsymbol\beta\\
\mathbf{X}\boldsymbol\beta
\end{pmatrix},
\begin{pmatrix}
\tilde{\mathbf{C}}_{\mathbf{X}^*} & \tilde{\mathbf{C}}_{\mathbf{X}^*,\mathbf{S}}\\
\tilde{\mathbf{C}}_{\mathbf{X}^*,\mathbf{S}} & \tilde{\mathbf{C}}_\mathbf{S}
\end{pmatrix}
\end{pmatrix}.
\end{align}
This implies directly that:
\begin{align}
\mathbf{y}^*
|\mathbf{y}_S,\mathbf{X}^*,\mathbf{X}, \boldsymbol\beta, \sigma^2, \gamma, \rho, \mathcal{A}
\sim
\mathcal{N}(
\mathbf{X}^*\boldsymbol\beta + \tilde{\mathbf{C}}_{\mathbf{X}^*,\mathbf{S}}\tilde{\mathbf{C}}_\mathbf{S}^{-1}(\mathbf{y}_S-\mathbf{X}\boldsymbol\beta), \tilde{\mathbf{C}}_{\mathbf{X}^*}-\tilde{\mathbf{C}}_{\mathbf{X}^*,\mathbf{S}}\tilde{\mathbf{C}}_\mathbf{S}^{-1}\tilde{\mathbf{C}}_{\mathbf{X}^*,\mathbf{S}}
)\label{pred1}
\end{align}
We use the expected value of the posterior predictive distribution as a prediction. Using Monte Carlo Integration, this quantity can easily be approximated as follows.
\begin{align}
\widehat{\mathbf{y}}^*&=\int \mathbf{y}^* p(\mathbf{y}^*|\mathbf{y}_S,\mathbf{X}^*,\mathbf{X})~ d\mathbf{y}^*, \nonumber\\
&=\int \mathbf{y}^*\left\{\int\int\int\int\int p(\mathbf{y}^*
|\mathbf{y}_S,\mathbf{X}^*,\mathbf{X},\boldsymbol\beta, \sigma^2, \gamma, \rho, \mathcal{A})p(\boldsymbol\beta, \sigma^2, \gamma, \rho, \mathcal{A}|\mathbf{y}_S,\mathbf{X}) \right. \nonumber \\ 
& \qquad \left. d\sigma^2 d\gamma d\rho\#(d\mathcal{A})(\lambda+\delta_0)^d(d\boldsymbol\beta)\right\}~ d\mathbf{y}^*,\nonumber\\
&=\int\int\int\int\int\left\{\int \mathbf{y}^* p(\mathbf{y}^*
|\mathbf{y}_S,\mathbf{X}^*,\mathbf{X},\boldsymbol\beta, \sigma^2, \gamma, \rho, \mathcal{A}) ~ d\mathbf{y}^*\right\}p(\boldsymbol\beta, \sigma^2, \gamma, \rho, \mathcal{A}|\mathbf{y}_S,\mathbf{X}) \nonumber \\
& \qquad {} d\sigma^2 d\gamma d\rho\#(d\mathcal{A})(\lambda+\delta_0)^d(d\boldsymbol\beta),\nonumber\\
&\underset{\eqref{pred1}}{=}\int\int\int\int\int\left\{   
\mathbf{X}^*\boldsymbol\beta + \tilde{\mathbf{C}}_{\mathbf{X}^*,\mathbf{S}}\tilde{\mathbf{C}}_\mathbf{S}^{-1}(\mathbf{y}_S-\mathbf{X}\boldsymbol\beta) \right\}p(\boldsymbol\beta, \sigma^2, \gamma, \rho, \mathcal{A}|\mathbf{y}_S,\mathbf{X}) \nonumber \\
& \qquad {} d\sigma^2 d\gamma d\rho\#(d\mathcal{A})(\lambda+\delta_0)^d(d\boldsymbol\beta),\nonumber\\
&\approx\frac{1}{N}\sum\limits_{i = 1}^N\mathbf{X}^*\boldsymbol\beta^{(i)} + \tilde{\mathbf{C}}_{\mathbf{X}^*,\mathbf{S}}^{(i)}\tilde{\mathbf{C}}_\mathbf{S}^{^{(i)}-1}(\mathbf{y}_S-\mathbf{X}\boldsymbol\beta^{(i)})\label{pred2}.
\end{align}
where $\#$ denotes the counting measure.
In  \eqref{pred2} notation $(i)$ shall point out that each of the individual summands is computed for an $iid$ sample from the posterior
$p(\boldsymbol\beta, \sigma^2, \gamma, \rho, \mathcal{A}|\mathbf{y}_S,\mathbf{X})$.

\section{Evaluation}
\label{sec:eval}
In this section, we evaluate and compare the predictive performance of our approach (VRNNGP) 
against several established feature selection methods. 
The performance comparison relies on accuracy measures: mean squared error (MSE) and mean absolute deviation (MAD).
These metrics are computed using testing datasets that are distinct from the training data, ensuring no overlap in observations between the two sets.
The methods being compared include:
\begin{itemize}
\item \textbf{Lasso.} The least absolute shrinkage and selection operator (Lasso) \citep{Tibshirani1996} penalizes the absolute values of the regression coefficients in the classical linear model in order to obtain a sparse solution. The R package \texttt{glmnet} \citep{Friedman2010} implements the lasso. Cross-validation is used to determine the penalization strength. 
\item \textbf{Adaptive Lasso (ALasso).} The adaptive Lasso \citep{Zou2006} is an extension of the classical lasso that allows for individual penalization strengths regarding the regression coefficients. In this comparison study, the individual penalization factors are defined by multiplying a global factor with the inverse numbers of the absolute values of the optimized regression coefficients from a ridge-penalized model. Again, the package \texttt{glmnet} is used for this task.
\item \textbf{Bayes Linear Selection (BLS).}  Posch et al.  \cite{POSCH2020} proposed a Bayesian variable selection approach for linear regression models. Since our approach shares some common features, e.g., the prior for the random set $\mathcal{A}$, we also include their method in our evaluation study.
\item \textbf{Random Forest (RF).} A combination of tree predictors proposed by Breiman  \cite{Breiman2001}. The method is implemented in the R package \texttt{randomForest} \citep{Liaw2002}.
\item \textbf{Robust Gaussian Process (RGP).} The R package \texttt{RobustGaSP} \citep{Gu2020} implements robust Gaussian process regression models using reference priors and separable (anisotropic) covariance functions. We use this package with the default specifications (i.e. a Mat{\'e}rn $5/2$ kernel etc.), except that we force the method to estimate a nugget term. Note that GP models with anisotropic kernels allow for an automatic relevance determination \citep{Rasmussen2005}. To this aim, the predictors have to be scaled prior to model inference.
\item \textbf{Maximum Likelihood Gaussian Process (MLGP).} The R package \texttt{mlegp} \citep{Garrett2008} implements Gaussian Process regression using maximum likelihood estimates and an anisotropic covariance function. We use this package with the default settings for the model inference, except that we specify the mean function to be non-constant.

\item 
\textbf{Bayesian Adaptive Sampling (BAS)}. The \texttt{BAS} R package uses
 Zellner's g-prior or mixtures of g-priors corresponding to the Zellner-Siow Cauchy Priors or the mixture of g-priors from   \cite{doi:10.1198/016214507000001337}  for the prior distributions on coefficients of linear models.

\item  \textbf{Variational Inference for Bayesian Variable Selection (VARBVS).} 
The \texttt{varbvs} R package 
uses simulations designed to mimic genetic association studies. 
Carbonetto and Stephens \cite{Carbonetto}  show that this simple variational approximation yields posterior inferences in some settings that closely match exact values.

\item \textbf{Bayesian Additive Regression Trees (BART).} 
The \texttt{BART} R package (\cite{bartpackage}) implements 
 a Bayesian nonparametric, machine learning, ensemble predictive modeling method.
 It is a tree-based method that fits the outcome to 
 an arbitrary random function of the covariates.

\item
\textbf{Bayesian Additive
Regression Trees Approach using Gaussian Processes (GP-BART).}
 The lack of smoothness and the absence of an explicit covariance structure over the observations in BART can yield poor performance in cases where such assumptions would be necessary.
 Maia et al.
 \cite{MAIA2024107858} propose a model which is an extension of BART which addresses this limitation by assuming Gaussian process priors for the predictions of each terminal node among all trees.
\end{itemize}

We consider four different benchmark data sets where the first three represent the common way of benchmarking variable selection methods:
the Pepelyshev function (Section \ref{ss:pepe}),
the  Sine function (Section  \ref{ss:sin}), and the 
  body fat data set (Section \ref {ss:body}).
Additionally, we consider a data set obtained during the semiconductor manufacturing process (Section \ref{ss:apc}).

In all experiments, the number of nearest neighbors is set to $m=10$, as preliminary experiments showed that this number is sufficiently large to obtain proper results.
 For all considered data sets, 
we assume that no predictor is preferred over another, i.e.,
the a priori importance of the predictors is set to $p_1=\ldots=p_d=1/d$  .
For data sets used in Sections \ref{ss:pepe}, \ref{ss:sin}, and  \ref{ss:body},
the  mapping $\tilde{p}(\cdot)$, which represents the a priori belief regarding the number of predictors contributing to the model,
is defined as $\tilde{p}(k)=1/k,~\forall k\in\{1,\ldots,d\}$. 
Thus, smaller models are quite intuitively preferred.
 The parameters $\tilde{p}_i$ are specified like this since there is no a priori knowledge available, and a specification using empirical investigations regarding the importance of the individual predictors is not necessary or possible.
 We observe that the MCMC chains converge without informative values for the $\tilde{p}_i$ after a reasonable amount of MCMC iterations (a few thousand). 

In contrast to that, for the semiconductor manufacturing data (Section \ref{ss:apc}) $\tilde{p}(\cdot)$ is defined to be proportional to the third power of the probability mass function of a zero truncated binomial distribution with size parameter $d$ and second parameter $1/d$, as proposed by \cite{POSCH2020}. This is necessary, as the data set has a larger number of potential predictors.

Regarding the model inference,
we set the tuning parameter $p_h$ to $0.6$, which means that 
 the proposal for $\mathcal{A}$ slightly prefers to change $\mathcal{A}$ over keeping it unchanged.
  For simplicity, nearly all tuning parameters of the HMC step are fixed to the values $m_{\tilde{\rho}}=m_{\tilde{\gamma}}=1,L=2$.
 The only exception is the step size $S$ of the leapfrog method. This parameter turns out to be the most influential one, and values between $0.15$ and $0.5$ are chosen. The actual specification is selected in such a way (per trial and error) that the MCMC chains show a good mixing.

\subsection{Pepelyshev Function}
\label{ss:pepe}

At first, we test the considered methods on the Pepelyshev function \citep{Dette2010}, a commonly used test function in computer experiments.
In this field, it has become common to use a GP in order to model the output of a complex system \citep{VOLLERT2019}.
The highly curved Pepelyshev function is given by
\begin{align}
f(\mathbf{x}) = 4(x_1-2+8x_2-8x_2^2)^2+(3-4x_2)^2+16\sqrt{x_3+1}(2x_3-1)^2, \label{Pepel}
\end{align}
for $x_1,x_2,x_3\in[0,1]$. 
By extending Function \eqref{Pepel} by $17$ inert variables $x_4,\ldots,x_{20}\in[0,1]$, 
we create a variable selection problem.

The evaluation of our approach is based on a training data set with a sample size of $31$ and a testing dataset of size $100$. Both datasets are constructed by generating a maxmin latin hypercube design (LHD), respectively. Maxmin LHDs are standard designs for computer experiments \cite{Santner2003}. They can be easily constructed by using the R package \texttt{lhs} \cite{Carnell2020}. Prior to the analysis, the training and the testing dataset are scaled. 

The results are reported in Table \ref{tab:ACCPepe}.
One can see that VRNNGP outperforms all other considered methods.
The superior performance is because the other methods fail to identify the active variables $x_1,x_2,x_3$, i.e., they also use the inert variables.
However, it should be mentioned that VRNNGP cannot properly identify $x_1,x_2,x_3$. In the VRNNGP model,
variables $x_2,x_3$ have a posterior probability of $1$ to contribute to the models, while all other predictors have a probability close to zero.

\begin{figure*}
\centering
\begin{minipage}{.48\textwidth}
  \centering
  \includegraphics[width=\linewidth]{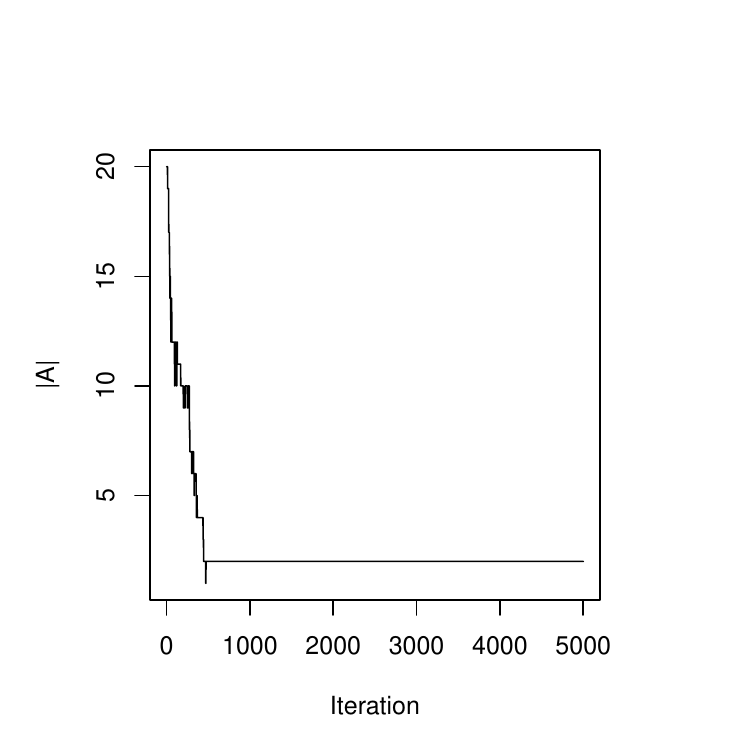}
  \caption{Markov chain of the cardinality of the random set $\mathcal{A}$ for the Pepelyshev Function with inert variables.}
  \label{fig:MC-k}
\end{minipage}%
\hspace{.01\textwidth}
\begin{minipage}{.48\textwidth}
  \centering
  \includegraphics[width=\linewidth]{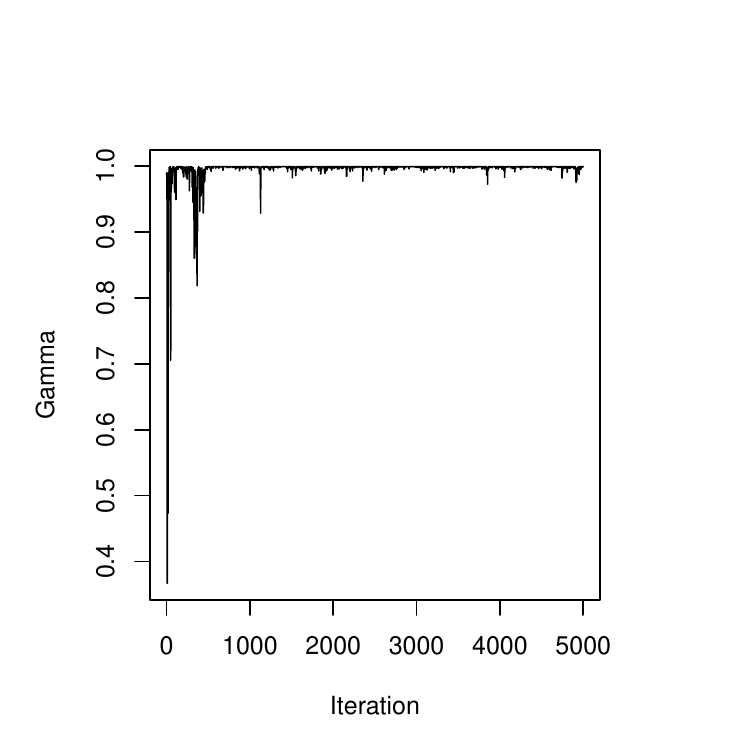}
  \caption{Markov chain of the covariance parameter $\gamma$ for the Pepelyshev Function with inert variables.}
  \label{fig:MC-gamma}
\end{minipage}  
  \begin{minipage}{.48\textwidth}
  \centering
  \includegraphics[width=\linewidth]{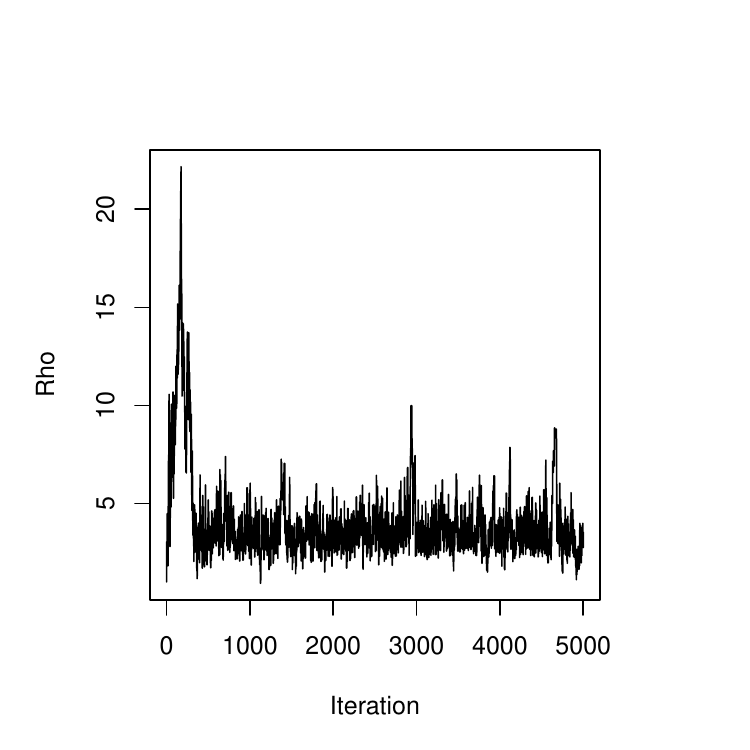}
  \caption{Markov chain of the covariance parameter $\rho$ for the Pepelyshev Function with inert variables.}
  \label{fig:MC-rho}
\end{minipage}%
\hspace{0.01\textwidth}
\begin{minipage}{.48\textwidth}
  \centering
  \includegraphics[width=\linewidth]{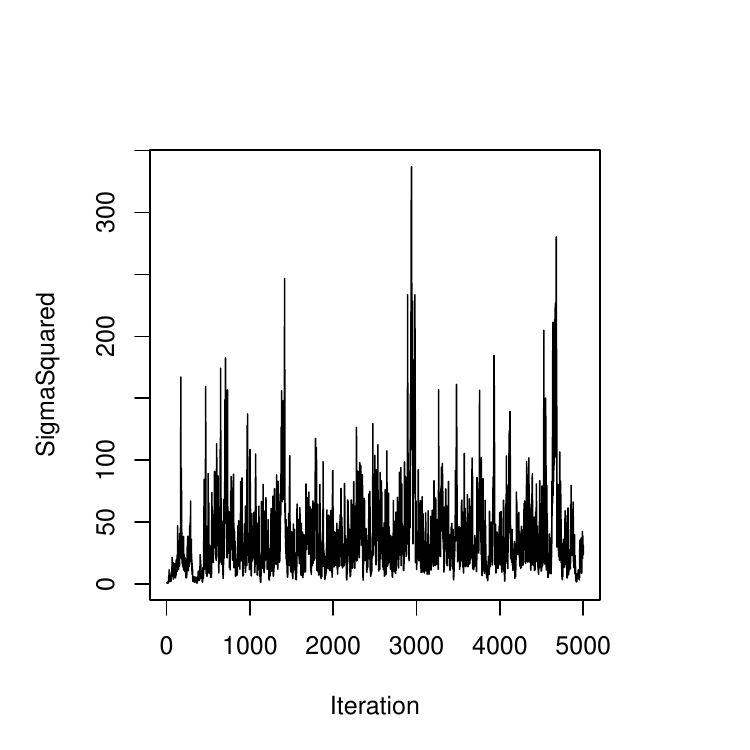}
  \caption{Markov chain of the covariance parameter $\sigma^2$ for the Pepelyshev Function with inert variables.}
  \label{fig:MC-sigma}
\end{minipage}    
\end{figure*}

Additionally, we want to illustrate that the proposed Metropolis-within-Gibbs algorithm (Section \ref{ss:inference}) converges fast.
Figures \ref{fig:MC-k}-\ref{fig:MC-sigma} show the Markov chains of some representative model parameters. 
It can be seen that the algorithm converges after about $1000$ iterations.

\begin{table}[!ht]
\caption{
 Accuracy measures reported for all considered approaches regarding the Pepelyshev function with inert variables.}
\label{tab:ACCPepe}     
\centering
\begin{tabular}{l S[table-format=1.4, round-mode = places,
round-precision = 4] S[table-format=1.4, round-mode = places,
round-precision = 4]}
\hline
Method & {MSE} & {MAD} \\\hline 
VRNNGP  &0.1949179       &0.3148653 \\
MLGP    &1.999522        &1.176881 \\
RGP     &1.483502   	 &1.01951 \\
RF      &0.7351364       &0.7156615 \\
Lasso   &1.003959   	 &0.8617405 \\
ALasso  &1.63588    	 &1.065925 \\
BLS     &1.30754    	 &0.972762 \\
 BART  &      1.42512 &    0.9803931  \\
 GPBART & 1.65591 & 1.035887\\
BAS &  0.9984872  & 0.8677115  \\
 VARBVS  &     1.01961    &  0.8740759\\
\hline
\end{tabular}
\end{table}

\subsection{Sine Function}
\label{ss:sin}

In a simulation study, adapted from Savitsky et al. \cite{Savitsky2011},
the observed continuous response $y$ is constructed from a nonlinear relationship of two variables ($x_1$ and $x_2$).
The  model contains 20 normally distributed variables; hence $x_3,x_4,\ldots,x_{20}$ are inert variables.
The full model from which the data is generated reads as follows:
\begin{align}
y &= \operatorname{sin}(x_1)+\operatorname{sin}(5x_2)+\varepsilon,
\nonumber\\
\varepsilon &\sim\mathcal{N}(0,0.0025),  \nonumber\\
(x_1,\ldots,x_{20})^T &\sim\mathcal{N}(\mathbf{0},\mathbf{\Sigma}),\nonumber\\
\operatorname{diag}(\mathbf{\Sigma})&=\mathbf{1}, \nonumber\\
\Sigma_{ij}&=\begin{cases}
0.4, &\mbox{ if } (i,j)\in\{(1,13),(13,1)\},\\
0.3, &\mbox{ if } (i,j)\in\{(1,12),(12,1)\},\\
0.4, &\mbox{ if } (i,j)\in\{(2,14),(14,2)\},\\
0.3, &\mbox{ if } (i,j)\in\{(2,15),(15,2)\},\\
0, &\mbox{ else.}
\end{cases}
\nonumber
\end{align}
 In particular, 100 samples are simulated from the model to obtain a dataset that is subsequently zero-centered and scaled.
 The performance evaluation is based on a $5$-fold cross-validation.
 The results are reported in Table \ref{tab:ACCsinus}.
 We observe that VRNNGP performs significantly better than the other methods. 
 Indeed, this method identifies the active variables $x_1,x_2$ in four of 
 the five considered train/test splits, while the other methods cannot do so.

\begin{table}[!ht]
\caption{
 Accuracy measures reported for all considered approaches regarding the Sine function simulation model.}
\label{tab:ACCsinus}     
\centering
\begin{tabular}{l S[table-format=1.4, round-mode = places,
round-precision = 4] S[table-format=1.4, round-mode = places,
round-precision = 4]}
Method & {MSE} & {MAD} \\\hline 
VRNNGP  &0.3580486       &0.4316482 \\
MLGP    &1.219693        &0.9032923 \\
RGP     &0.7396772   	 &0.7100051 \\
RF      &0.7706562       &0.7031102 \\
Lasso   &0.8173408   	 &0.7418092 \\
ALasso  &0.8286106    	 &0.7483805 \\
BLS     &0.8528274    	 &0.7460358 \\
BART  &  1.449045    &  0.9566844  \\
GP-BART & 1.484425  &  0.9744137 \\
BAS  &  0.7727341 & 0.726057  \\
VARBVS  &   0.7747144      &  0.7295798 \\
\hline
\end{tabular}
\end{table}

\subsection{Body Fat Data Set}
\label{ss:body}
The body fat data set \citep{Johnson1996}  includes 128 observations on 14 variables (available in the \texttt{mplot} R package \citep{Tarr2018}).
The target variable is the body fat percentage of a male, and the predictors are age, weight, height, and ten body circumference measurements. 
A 5-fold cross-validation is performed to evaluate and compare the performance of VRNNGP against other methods. 
Prior to the analysis, the whole dataset was zero-centered and scaled.
The results are reported in Table \ref{tab:ACCbody}.

\begin{figure}
\centering
  \includegraphics[width=0.8\linewidth]{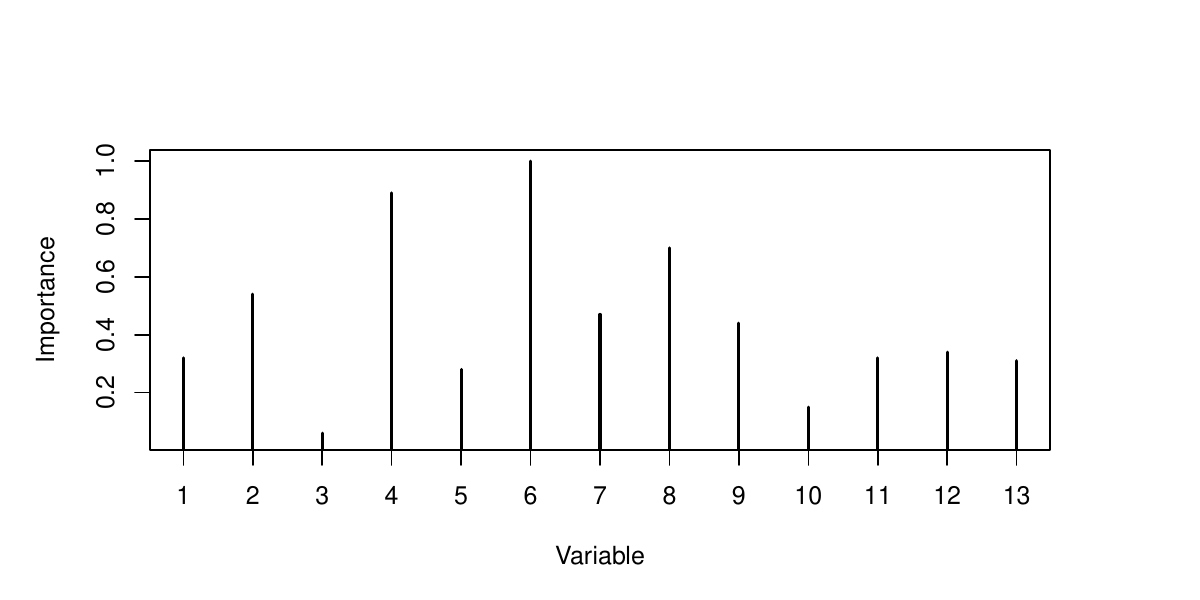}
  \caption{Variable Importance obtained from VRNNGP on the body fat data set.}
  \label{fig:imp-body-NNGP}
\end{figure}

\begin{figure}
\centering
  \includegraphics[width=0.8\linewidth]{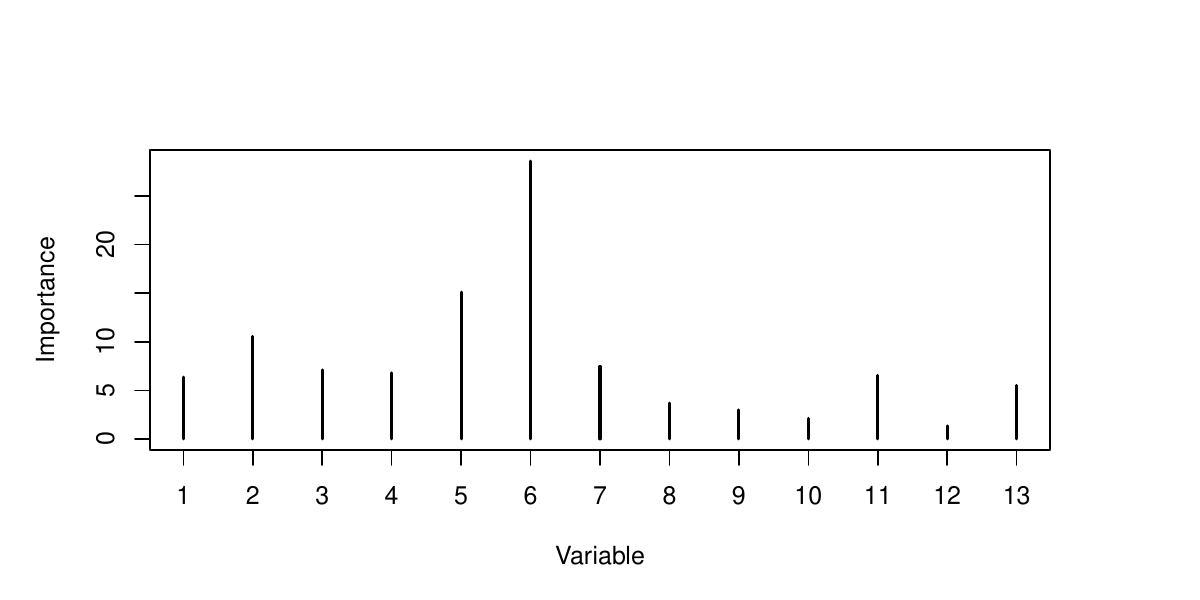}
  \caption{Variable Importance obtained from RF on the body fat data set.}
  \label{fig:imp-body-RF}
\end{figure}

\begin{figure}
\centering
    \includegraphics[width=0.8\linewidth]{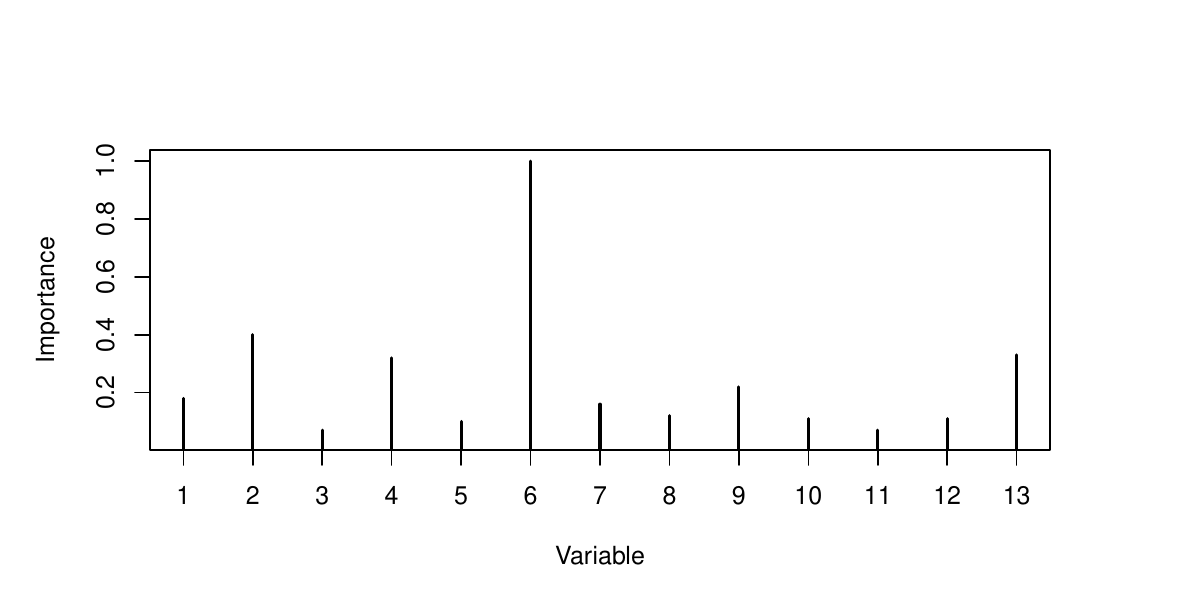}
  \caption{Variable Importance obtained from BLS on the body fat data set.}
  \label{fig:imp-body-BLS}
\end{figure}

We observe that VRNNGP achieves the best prediction results. The BAS library achieves nearly the same prediction quality.
In Figures \ref{fig:imp-body-NNGP}-\ref{fig:imp-body-BLS}, the importance of the individual predictors (compare to Section \ref{ss:apc}) obtained from the methods VRNNGP, RF, and BLS are visualized.
While all of these methods agree that the sixth predictor (abdomen circumference) is the most influential, there are some disagreements regarding the order of importance of the other predictors.

\begin{table}[!ht]
\caption{Accuracy measures reported for all considered approaches regarding the body fat data set.}
\label{tab:ACCbody}   
\centering
\begin{tabular}{l S[table-format=1.4, round-mode = places,
round-precision = 4] S[table-format=1.4, round-mode = places,
round-precision = 4]}
Method & {MSE} & {MAD} \\\hline 
VRNNGP  & \bf 0.2762897        & \bf 0.4211418\\
MLGP    &0.3657517        & 0.4865936\\
RGP     &0.308341   	  & 0.4434407\\
RF      &0.3035638        & 0.4458874\\
Lasso   &0.3113187   	  & 0.4484568\\
ALasso  &0.314493    	  & 0.4521927\\
BLS     &0.3014364    	  & 0.4356817\\
BART  &   1.743591   &   1.071021 \\
GP-BART& 0.5994538 & 0.620887 \\
BAS & 0.293194  &    0.4293307\\
VARBVS  &      0.3224403     & 0.4581236\\
\hline
\end{tabular}
\end{table}

\subsection{Semiconductor Frontend Manufacturing Data}
\label{ss:apc}

Semiconductor production involves a multitude of intricate steps, commencing with the frontend phase.
Semiconductor manufacturing is renowned for their high-tech setups, characterized by intricate processes, extensive automation, and significant levels of digitalization (\cite{Waschneck2017,Lee2023}).
Following the frontend phase, wafers progress to an interim storage facility. 
Subsequently, they move into the second and final production phase, known as the backend.
In this stage, wafers undergo processes where they are divided into individual chips, affixed onto a lead frame, and ultimately packaged and sealed.

The frontend production process involves transforming semiconductor material slices (wafers) through hundreds of steps.
Throughout each step, various sensors on the equipment record physical parameters—such as gas flow, voltage,  temperature, etc.
Some key numbers are derived from these records. These are usually statistical quantities like the recorded data's mean, standard deviation, or maximum values.
These key numbers serve to monitor process stability.
Once frontend production is completed, each wafer undergoes thorough testing, resulting in wafer test data. 
These tests, employing procedures that measure quantities like current, voltage, or resistance, aim to ensure each device meets its specifications.
If specific parameters from the wafer test data stand out, potentially indicating high failure rates, the source of these issues must be traced back to the frontend production stage.
Variable selection models help pinpoint specific production parameters that significantly impact the quality of the final product. 
We apply our approach to a data set collected in semiconductor frontend production \citep{Pleschberger2021}
which comprises 228 observations derived from 77 potential predictors (key numbers). 
These predictors are extracted from data recorded during two selected process steps. These two steps have been found to be most critical for the final result.
The target variable represents an analog measurement from the wafer test data, reflecting the quality of the wafer. Notably, higher measured values correspond to poorer quality.
Before conducting the analysis, we standardized the predictors and centered the target variable around zero. 
Additionally, the entire dataset was randomly split into two equal-sized subsets: a training dataset and a testing dataset. 
This division ensures an unbiased evaluation and validation of the model's performance.

\begin{figure}
\centering
  \includegraphics[width=0.8\linewidth]{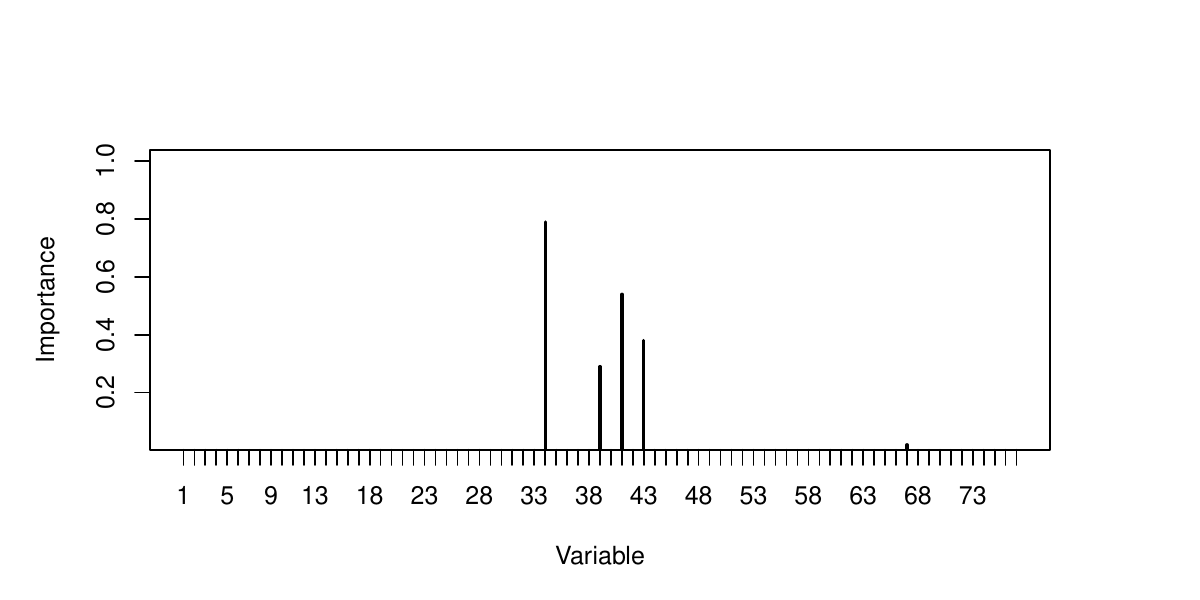}
  \caption{Variable Importance obtained from VRNNGP on the semiconductor frontend manufacturing data.}
  \label{fig:imp-NNGP}
  \end{figure}

\begin{figure}
\centering
  \includegraphics[width=0.8\linewidth]{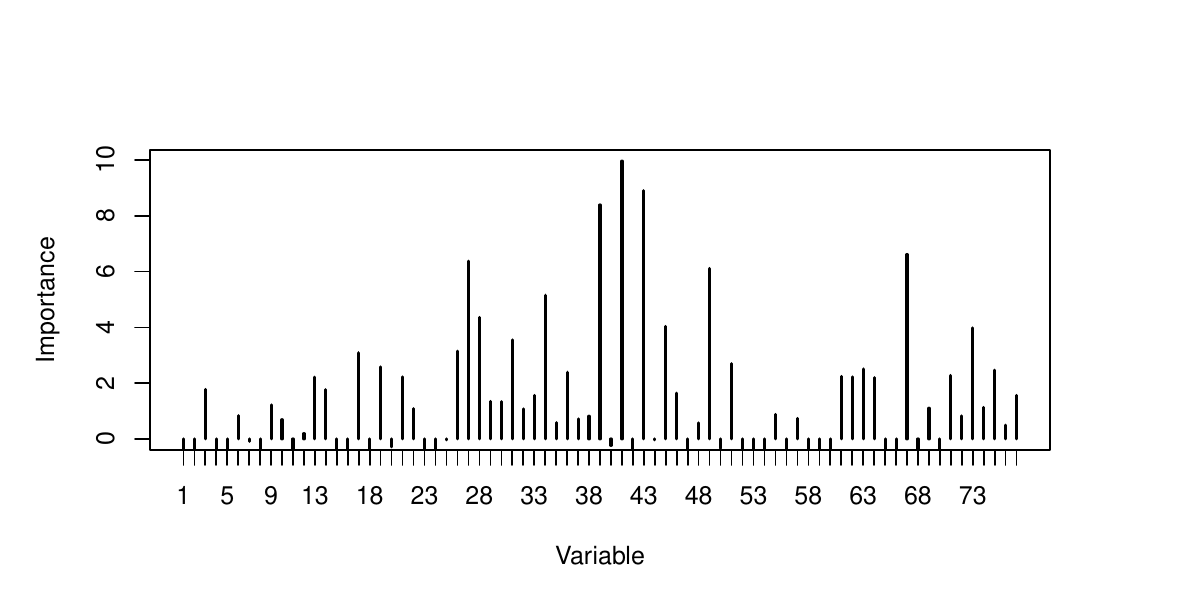}
  \caption{Variable Importance obtained from RF on the semiconductor frontend manufacturing data.}
  \label{fig:imp-RF}
\end{figure}

\begin{figure}
\centering
    \includegraphics[width=0.8\linewidth]{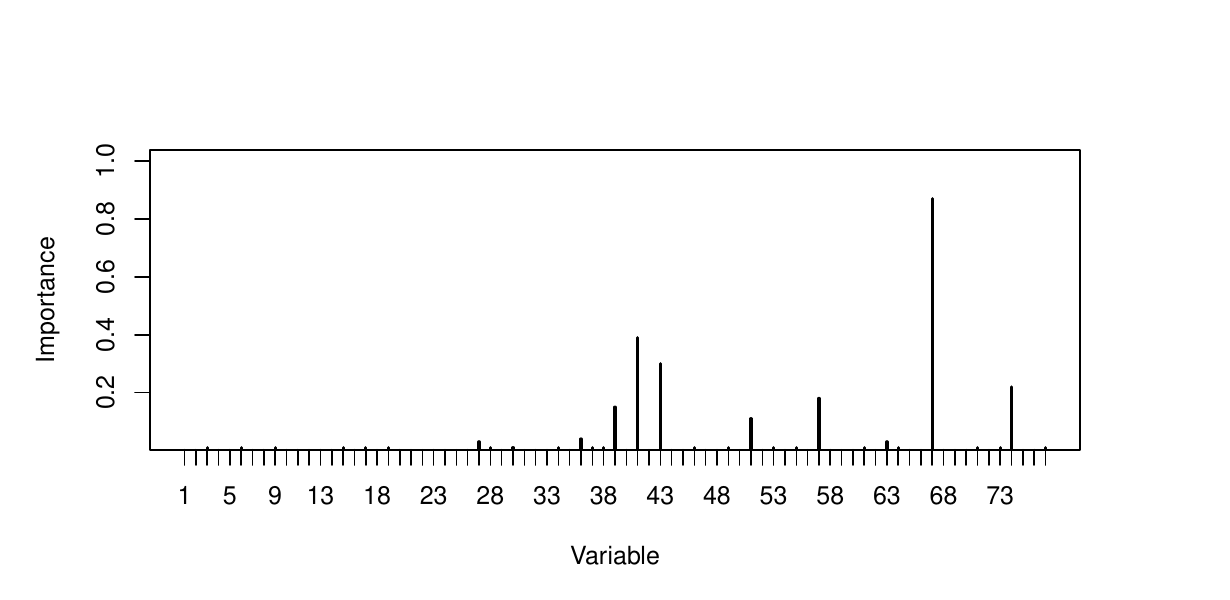}
  \caption{Variable Importance obtained from BLS on the semiconductor frontend manufacturing data.}
  \label{fig:imp-BLS}
\end{figure}

\begin{figure}
\centering
  \includegraphics[width=0.8\linewidth]{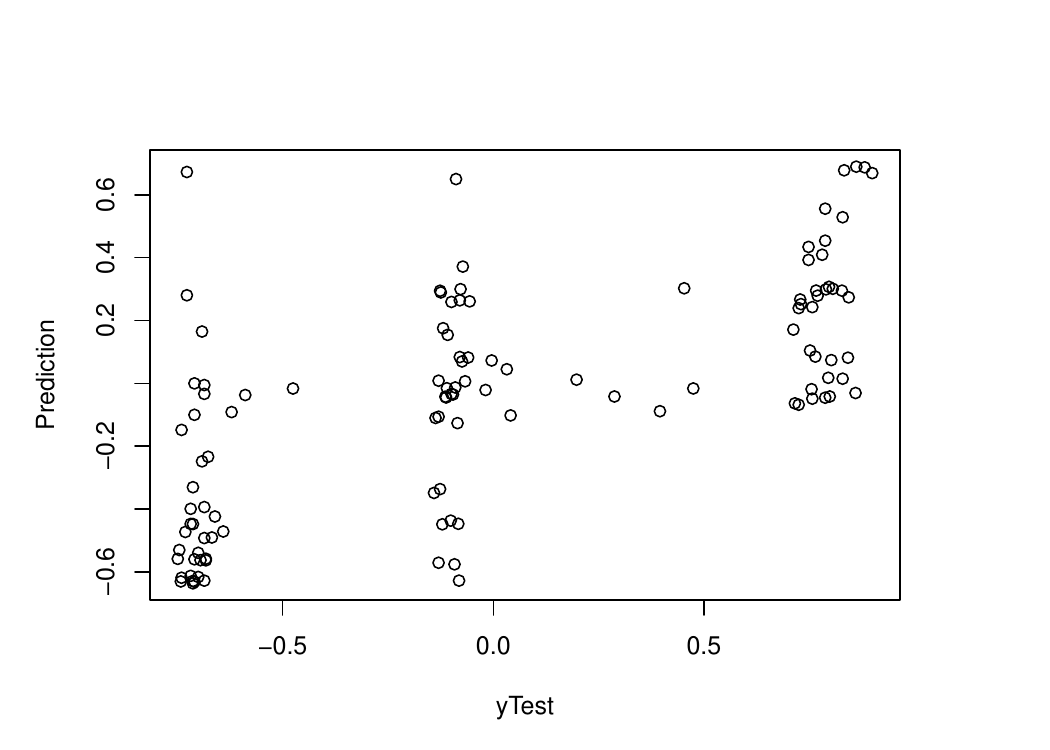}
  \caption{True production quality of the testing dataset (yTest) vs. predicted quality obtained by RGP on the  reduced semiconductor frontend manufacturing data.}
  \label{fig:RGP-pred}
\end{figure}

In Table \ref{tab:ACCAPC} the accuracy of the methods considered in this study is shown. VRNNGP returns the best results, followed by RF. MLGP fails completely, while the robust version of RGP performs slightly worse than the linear approaches. The Figures \ref{fig:imp-NNGP}-\ref{fig:imp-BLS} visualize the importance of the individual predictors obtained from the methods VRNNGP, RF, and BLS. For VRNNGP and BLS, the importance of a predictor is the relative frequency of how often its index lies in the posterior samples from $\mathcal{A}$. This can be interpreted as the posterior probability of the predictor being included in the model. For RF, the visualized importance is the mean decrease in accuracy if a particular variable is removed. This quantity can be easily computed using the \texttt{importance()} function included in the R package \texttt{randomForest}. It can be seen that the importance values differ from method to method. However, the two most important variables in NNGP (variable $34$, variable $41$) are also important in RF. In BLS, at least the variable $41$ is considered important. We assume that the influence of the variable $34$ is non-linear, such that a linear model fails to detect this influence. Since the variables $34$ and $41$ appear to be important across models, we train the two models, RGP and RF, again, but with a reduced dataset including only these two variables. In Table \ref{tab:ACCAPC2}, the respective accuracies are presented.
Both models perform similarly on the reduced dataset but significantly better than before on the complete dataset.
 In particular, the reduction to the variables $34$ and $41$ led to the best results overall. The Figures show the functions learned by RGP and RF based on the reduced dataset. Both models agree on parameter specifications with poor production quality. Finally, in Figure \ref{fig:RGP-pred}, the true production quality of the testing dataset(yTest) is plotted vs. the predicted quality from the RGP model with the reduced dataset. It can be observed that all of the really poor production results (yTest $>$ $0.5$) can be identified. This shows that the variables selected by NNGP are useful for this application.

\begin{table}[!ht]
\caption{Accuracy measures reported for all considered approaches regarding the semiconductor frontend manufacturing data.}
\label{tab:ACCAPC}     
\centering
\begin{tabular}{l S[table-format=1.4, round-mode = places,
round-precision = 4] S[table-format=1.4, round-mode = places,
round-precision = 4]}
Method & {MSE} & {MAD} \\\hline 
VRNNGP  & \bf 0.2389607       & \bf0.4197644 \\
MLGP      &32.49251        & 1.44142\\
RGP     &0.3108199   	 &0.4695633 \\
RF      &0.2687667       &0.4441394 \\
Lasso   &0.2843513   	 &0.4568827 \\
ALasso  &0.2903017     	 &0.4529153 \\
BLS     &0.2846291   	 &0.4471566 \\
BART  &   0.4486588   &   0.570761 \\
GP-BART&  0.5234023 & 0.5741475 \\
BAS & 0.2993064  &   0.4617779\\
VARBVS  &       0.3028562    &  0.4601329\\
\hline
\end{tabular}
\end{table}

\begin{table}[!ht]
\caption{
Accuracy measures reported for RGP and RF regarding reduced semiconductor frontend manufacturing data where only the variables $34$ and $41$ are considered.}
\label{tab:ACCAPC2}    
\centering
\begin{tabular}{l S[table-format=1.4, round-mode = places,
round-precision = 4] S[table-format=1.4, round-mode = places,
round-precision = 4]}
Method & {MSE} & {MAD} \\\hline 
RGP     &0.2053222  	    & 0.3692166\\
RF      &0.2022327          &0.3539577 \\
\hline
\end{tabular}
\end{table}

\begin{figure}
\centering
  \includegraphics[width=0.8\linewidth]{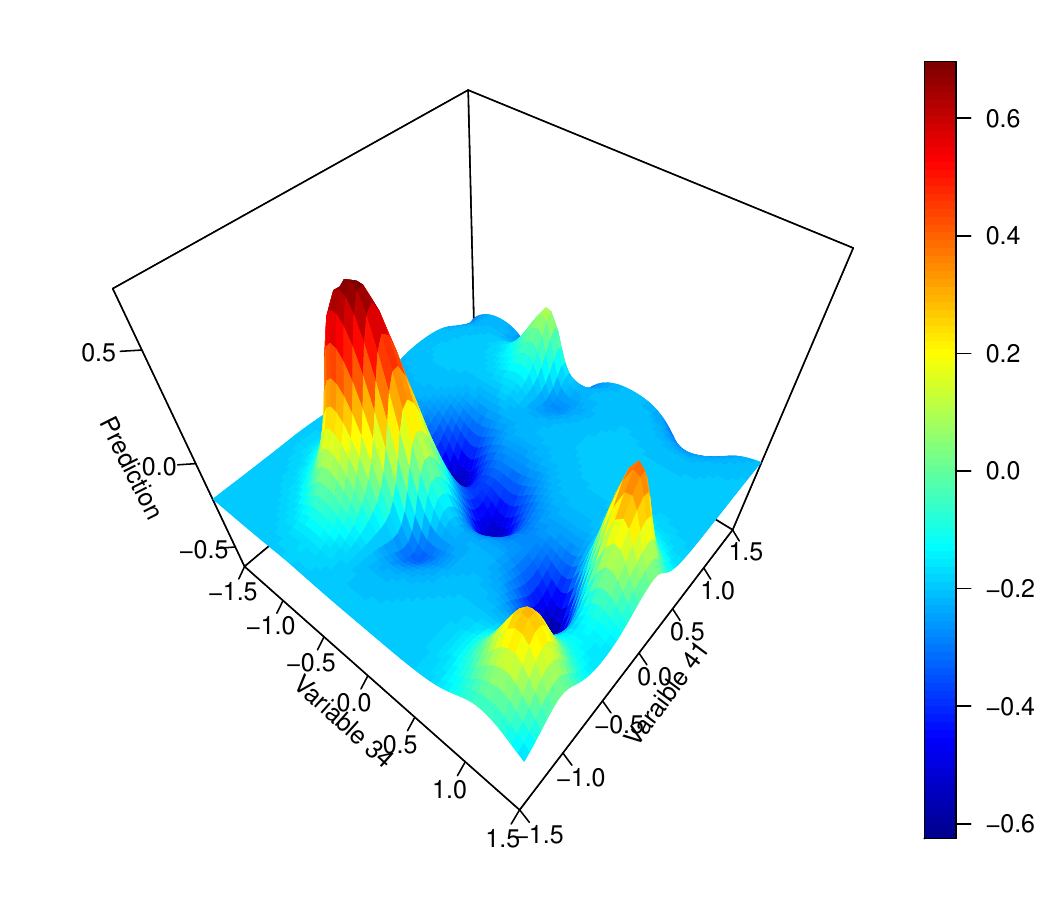}
  \caption{Model fit obtained from RGP on the semiconductor frontend manufacturing data.}
  \label{fig:fit-RGP}
  \end{figure}
  
\begin{figure}
\centering  \includegraphics[width=0.8\linewidth]{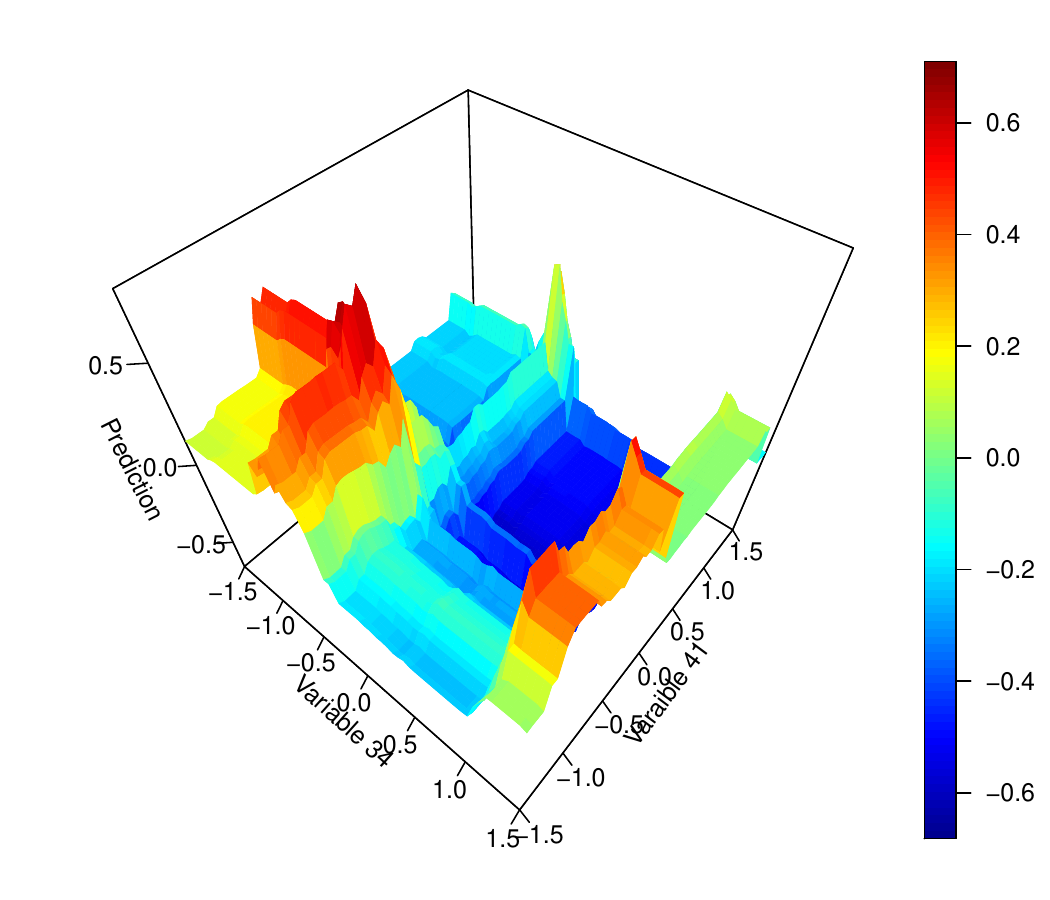}
  \caption{Model fit obtained from RF on the semiconductor frontend manufacturing data.}
  \label{fig:fit-RF}
\end{figure}

\section{Conclusion}
\label{sec:conclusion}
In this study, we introduced a novel Bayesian approach to solve the variable selection problem.
We combined nearest neighbor Gaussian processes, reference priors, and model selection strategies based on a random set to this aim.
To the best of our knowledge, this specific combination has not been investigated in prior research.

Using nearest neighbor processes reduces computational costs compared to traditional stochastic processes, while employing reference priors ensures numerical robustness in model inference. 
The random set approach allows the selection of the most predictive variables while incorporating relevant a priori knowledge. 
For model inference, we proposed a Metropolis-within-Gibbs algorithm.
The algorithm allows for transitions between different sets of active predictors (predictors that truly contribute to the model) and further uses Hamiltonian dynamics for the inference of the Gaussian process correlation parameters. 

In a computational study, the prediction quality of the proposed method was evaluated.
We find that our approach could identify truly significant variables in evaluations, including simulated data and an approximation problem corresponding to computer experiments.
Linear models, anisotropic Gaussian processes (automatic relevance determination), and random forest models fell short in these variable selection tasks.
Finally, the results for a variable selection problem that occurs in semiconductor manufacturing
highlight the potential contribution of our proposed method in high-tech manufacturing applications.

\section*{Acknowledgements}
The publication is an outcome stemming from the iDev40 project  (\url{www.idev40.eu}). 
The iDev40 project is co-funded by the consortium members and ECSEL Joint Undertaking under grant agreement No 783163. 
The JU receives support from the European Union's Horizon 2020 research and innovation programme, national grants from Austria, Belgium, Germany, Italy, Spain and Romania as well as the European Structural and Investement Funds.
The information and results set out in this publication are those of the authors and do not necessarily reflect the opinion of the ECSEL Joint Undertaking.

\section*{Declarations}

\subsection*{Competing interests.}
The authors have no competing interests to declare that are relevant to the content of this article.

\subsection*{Author contribution.}
The methodology was proposed by Konstantin Posch and J\"urgen Pilz.
Konstantin Posch and Maximilian Arbeiter did the implementation work.
The domain-specific data set was curated by Martin Pleschberger.
Computational experiments were performed by Konstantin Posch and Christian Truden.
The first draft of the manuscript was written by Konstantin Posch, and Christian Truden revised it. J\"urgen  Pilz supervised the process leading to this manuscript at all times.
All authors read and approved the final manuscript.

\appendix

\section{Computation of Derivatives}\label{secA1}

The computation of the derivatives presented in Equation \eqref{deriv1} and Equation \eqref{deriv2} is given as follows.

\begin{align}
\frac{\partial \mathbf{AB}}{\partial t}&=\frac{\partial \mathbf{A}}{\partial t}\mathbf{B}+\mathbf{A}\frac{\partial\mathbf{B}}{\partial t}, \nonumber\\
\frac{\partial\mathbf{A}^{-1}}{\partial t}&=-\mathbf{A}^{-1}\frac{\partial\mathbf{A}}{\partial t}\mathbf{A}^{-1} , \nonumber \\
\frac{\partial\boldsymbol{\tilde{\mathcal{K}}}_{S}}{\partial \rho}&=\frac{\partial}{\partial\rho}\mathbf{B}_{S}^{-1}\underbrace{\operatorname{diag}(\mathcal{F}_{\mathbf{x}_1},\ldots,\mathcal{F}_{\mathbf{x}_n})}_{=:\boldsymbol{\mathcal{F}}_S}(\mathbf{B}_{S}^T)^{-1}, \nonumber\\
&=\frac{\partial\mathbf{B}_{S}^{-1}\boldsymbol{\mathcal{F}}_S}{\partial\rho}(\mathbf{B}_{S}^T)^{-1}+\mathbf{B}_{S}^{-1}\boldsymbol{\mathcal{F}}_S\frac{\partial(\mathbf{B}_{S}^T)^{-1}}{\partial\rho}, \nonumber\\
&=\left(\frac{\partial\mathbf{B}_{S}^{-1}}{\partial\rho}\boldsymbol{\mathcal{F}}_S+\mathbf{B}_{S}^{-1}\frac{\partial\boldsymbol{\mathcal{F}}_S}{\partial\rho}\right)(\mathbf{B}_{S}^T)^{-1}+\mathbf{B}_{S}^{-1}\boldsymbol{\mathcal{F}}_S\frac{\partial(\mathbf{B}_{S}^T)^{-1}}{\partial\rho}, \nonumber\\
&=\left(-\mathbf{B}_{S}^{-1}\frac{\partial\mathbf{B}_{S}}{\partial\rho}\mathbf{B}_{S}^{-1}\boldsymbol{\mathcal{F}}_S+\mathbf{B}_{S}^{-1}\frac{\partial\boldsymbol{\mathcal{F}}_S}{\partial\rho}\right)(\mathbf{B}_{S}^T)^{-1}-\mathbf{B}_{S}^{-1}\boldsymbol{\mathcal{F}}_S(\mathbf{B}_{S}^T)^{-1}\frac{\partial(\mathbf{B}_{S}^T)}{\partial\rho}(\mathbf{B}_{S}^T)^{-1}, \nonumber\\
&=-\mathbf{B}_{S}^{-1}\left\{\underbrace{\frac{\partial\mathbf{B}_{S}}{\partial\rho}\mathbf{B}_{S}^{-1}\boldsymbol{\mathcal{F}}_S}_{=:\mathbf{A}}-\frac{\partial\boldsymbol{\mathcal{F}}_S}{\partial\rho}+\underbrace{\boldsymbol{\mathcal{F}}_S(\mathbf{B}_{S}^T)^{-1}\frac{\partial(\mathbf{B}_{S}^T)}{\partial\rho}}_{\mathbf{A}^T}\right\}(\mathbf{B}_{S}^T)^{-1}, \nonumber\\
&=-\mathbf{B}_{S}^{-1}\left\{\mathbf{A}+\mathbf{A}^T-\frac{\partial\boldsymbol{\mathcal{F}}_S}{\partial\rho}\right\}(\mathbf{B}_{S}^T)^{-1}, \nonumber
\\
\frac{\partial \mathbf{B}_{S}}{\partial \rho}&=\frac{\partial[B_{\mathbf{x}_i,j}]_{i,j=1,\ldots,n}}{\partial\rho}, \qquad \text{where } B_{\mathbf{x}_i,j}=\begin{cases}
1, & \text{if } i=j,\\
-\mathbf{B}_{\mathbf{x}_i}[l], & \text{if } \mathbf{x}_j = \mathbf{x}_i^l\text{ for some $l$}\\
0, & \text{else},
\end{cases}, \nonumber\\
\frac{\partial \mathbf{B}_{\mathbf{x}_i}}{\partial\rho}&=\frac{\partial \boldsymbol{\mathcal{K}}_{\mathbf{x}_i,N(\mathbf{x}_i)}\boldsymbol{\mathcal{K}}_{N(\mathbf{x}_i)}^{-1}}{\partial\rho}, \nonumber\\
&=\frac{\partial \boldsymbol{\mathcal{K}}_{\mathbf{x}_i,N(\mathbf{x}_i)}}{\partial\rho}\boldsymbol{\mathcal{K}}_{N(\mathbf{x}_i)}^{-1}+\boldsymbol{\mathcal{K}}_{\mathbf{x}_i,N(\mathbf{x}_i)}\frac{\partial \boldsymbol{\mathcal{K}}_{N(\mathbf{x}_i)}^{-1}}{\partial\rho}, \nonumber\\
&=\frac{\partial \boldsymbol{\mathcal{K}}_{\mathbf{x}_i,N(\mathbf{x}_i)}}{\partial\rho}\boldsymbol{\mathcal{K}}_{N(\mathbf{x}_i)}^{-1}-\boldsymbol{\mathcal{K}}_{\mathbf{x}_i,N(\mathbf{x}_i)} \boldsymbol{\mathcal{K}}_{N(\mathbf{x}_i)}^{-1}\frac{\partial \boldsymbol{\mathcal{K}}_{N(\mathbf{x}_i)}}{\partial\rho} \boldsymbol{\mathcal{K}}_{N(\mathbf{x}_i)}^{-1}, \nonumber\\
&=\left(\frac{\partial \boldsymbol{\mathcal{K}}_{\mathbf{x}_i,N(\mathbf{x}_i)}}{\partial\rho}-\mathbf{B}_{\mathbf{x}_i}\frac{\partial \boldsymbol{\mathcal{K}}_{N(\mathbf{x}_i)}}{\partial\rho}\right)\boldsymbol{\mathcal{K}}_{N(\mathbf{x}_i)}^{-1}. \nonumber
\intertext{This gives:}
\frac{\partial \mathbf{B}_{S}}{\partial \rho}&=
\begin{cases}
0, & \text{if}~ i\geq j,\\
\left\{\left(\mathbf{B}_{\mathbf{x}_i}\frac{\partial \boldsymbol{\mathcal{K}}_{N(\mathbf{x}_i)}}{\partial\rho}-\frac{\partial \boldsymbol{\mathcal{K}}_{\mathbf{x}_i,N(\mathbf{x}_i)}}{\partial\rho}\right)\boldsymbol{\mathcal{K}}_{N(\mathbf{x}_i)}^{-1}\right\}[l], & \text{if}~ i<j\text{ and }\mathbf{x}_j=\mathbf{x}_i^{l}\text{ for some }l , \\
0,&\text{else},
\end{cases}
\nonumber \\
\frac{\partial \boldsymbol{\mathcal{F}}_S}{\partial\rho}&=\operatorname{diag}\left(\frac{\partial\mathcal{F}_{\mathbf{x}_1}}{\rho},\ldots,\frac{\partial\mathcal{F}_{\mathbf{x}_n}}{\rho}\right) , \nonumber
\\
\frac{\partial\mathcal{F}_{\mathbf{x}_i}}{\rho}&=\frac{\partial \mathcal{K}_{\mathbf{x}_i}-\mathbf{B}_{\mathbf{x}_i}\boldsymbol{\mathcal{K}}_{N(\mathbf{x}_i),\mathbf{x}_i}}{\partial\rho}, \nonumber\\
&=\frac{\partial\mathcal{K}_{\mathbf{x}_i}}{\partial\rho}-\left(\frac{\partial\mathbf{B}_{\mathbf{x}_i}}{\partial\rho}\boldsymbol{\mathcal{K}}_{N(\mathbf{x}_i),\mathbf{x}_i}+\mathbf{B}_{\mathbf{x}_i}\frac{\partial\boldsymbol{\mathcal{K}}_{N(\mathbf{x}_i),\mathbf{x}_i}}{\partial\rho}\right)
, \nonumber
\end{align}
It is noteworthy that all the above calculations also hold true
 for $\mathbf{W}_{\gamma}$ when replacing $\rho$ with $\gamma$ such that:
\begin{align}
\frac{\partial\mathcal{K}_{\mathbf{x}_i,\mathbf{x}_j}}{\partial\rho}&= \gamma\frac{\partial K_{\mathbf{x}_i,\mathbf{x}_j}}{\partial\rho}, \nonumber\\
\frac{\partial\mathcal{K}_{\mathbf{x}_i,\mathbf{x}_j}}{\partial\gamma}&= -\delta_{(\mathbf{x}_i=\mathbf{x}_j)}+K_{\mathbf{x}_i,\mathbf{x}_j}, \nonumber \\
\frac{\partial K_{\mathbf{x}_i,\mathbf{x}_j}}{\partial\rho}&=\left[-\frac{\sqrt{5}d_{\mathcal{A}}}{\rho^2}-\frac{10d_{\mathcal{A}}^2}{3\rho^3}\right]\operatorname{exp}\left(-\frac{\sqrt{5}d_{\mathcal{A}}}{\rho}\right)+K_{\mathbf{x}_i,\mathbf{x}_j}\frac{\sqrt{5}d_{\mathcal{A}}}{\rho^2}. \nonumber
\end{align}

\end{document}